%% file: FF_in_SGF.tex
\documentclass[a4paper,11pt]{article}
\usepackage{}
\pdfoutput=1 

\usepackage{jheppub} 

\usepackage[T1]{fontenc} 

\usepackage{graphicx}
\graphicspath{{figures/}{fig/}}
\DeclareGraphicsExtensions{.eps}
\usepackage{amsmath}
\usepackage{bm}
\usepackage{slashed}
\usepackage{epsfig}
\usepackage{amsfonts}
\usepackage{epstopdf}
\usepackage{color}
\usepackage{extarrows}
\usepackage{multirow}
\usepackage[utf8]{inputenc}
\allowdisplaybreaks
\newcommand{\ben}{\begin{eqnarray}}
\newcommand{\een}{\end{eqnarray}}

\newcommand{\bef}{\begin{figure}[!htp]}
\newcommand{\eef}{\end{figure}}

\newcommand{\bea}{\begin{eqnarray}}
\newcommand{\eea}{\end{eqnarray}}

\def\ba{\begin{linenomath*}\begin{equation}}
\def\ea{\end{equation}\end{linenomath*}}

\allowdisplaybreaks

\newcommand{\state}[4]{{^{#1}\hspace{-0.6mm}#2_{#3}^{[#4]}}}
\newcommand{\stateprime}[4]{{^{#1}\hspace{-0.6mm}#2_{#3}^{\prime[#4]}}}

\newcommand\COcSa{\state{3}{S}{1}{8}}

\newcommand\lrd{\overleftrightarrow{D}}

\newcommand{\sect}[1]{\section{#1}}

\title{Fragmentation function of $g\to Q\bar{Q}(^3S_1^{[8]})$  in soft gluon factorization and threshold resummation}

\author[a]{An-Ping Chen}
\author[b]{Xiao-Bo Jin}
\author[b,c,d]{Yan-Qing Ma}
\author[b]{Ce Meng}

\affiliation[a]{College of Physics and Communication Electronics, Jiangxi Normal University, Nanchang 330022, China}
\affiliation[b]{School of Physics and State Key Laboratory of Nuclear Physics and
	Technology, Peking University, Beijing 100871, China}
\affiliation[c]{Center for High Energy physics, Peking University, Beijing 100871, China}
\affiliation[d]{Collaborative Innovation Center of Quantum Matter,
	Beijing 100871, China}

\emailAdd{chenanping@pku.edu.cn}
\emailAdd{xiaobojin@pku.edu.cn}
\emailAdd{yqma@pku.edu.cn}
\emailAdd{mengce75@pku.edu.cn}

\abstract{
We study the fragmentation function of the gluon to color-octet $^3S_1$ heavy quark-antiquark pair using the soft gluon
factorization (SGF) approach, which expresses the fragmentation function in a form of
perturbative short-distance hard part convoluted with one-dimensional color-octet $^3S_1$ soft gluon distribution (SGD). The short distance hard part is calculated to next-to-leading order in $\alpha_s$ and a renormalization group equation for the SGD is derived. By solving the renormalization group equation, threshold logarithms are resummed to all orders in perturbation theory. The comparison with gluon fragmentation function calculated in NRQCD factorization approach indicates that the SGF formula resums a series of velocity corrections in NRQCD which are important for phenomenological study.
}

\begin{document}

\maketitle

\sect{Introduction}	
\label{sec:intro}

Heavy quarkonium physics has been on the focus of much experimental and theoretical attention
since the discovery of the $J/\psi$ in 1974. As the simplest bound state of strong
interactions, heavy quarkonium is an ideal system to
study both perturbative and nonperturbative aspects of QCD. Our current understanding of
the decay and production of quarkonium is mainly based on the non-relativistic QCD (NRQCD)
factorization~\cite{Bodwin:1994jh}, which factorizes physical quantities into perturbatively calculable short-distance coefficients (SDCs) multiplied
by nonperturbative long-distance matrix elements (LDMEs).

However, recent studies shown that NRQCD factorization encounters some difficulties
in describing inclusive quarkonium production data. In Ref.~\cite{Ma:2017xno}, it was
argued that these difficulties may come from the bad convergence of velocity expansion in
NRQCD. The velocity expansion suffers from large high
order relativistic corrections due to ignoring the effects of soft hadrons emitted in the
hadronization process. For this reason, the authors proposed a new factorization approach,
called soft gluon factorization (SGF), to describe quarkonium production and decay~\cite{Ma:2017xno}. It was argued in Ref.~\cite{Chen:2020yeg} that the SGF is equivalent to the NRQCD factorization, but with a series of important relativistic corrections originated from kinematic effects resummed. As a result, the SGF approach should has a much better convergence in the velocity expansion. Indeed, it was found in Ref.~\cite{Ma:2017xno} that the lowest order result in velocity expansion in NRQCD factorization can deviate from
the full SGF result by a factor of 4, but the lowest order velocity expansion in SGF can deviate from
its full result by only a small value.

In the SGF approach, the quarkonium production cross section can be expressed in the following
factorization formula
\begin{align}\label{eq:fac4d}
 (2\pi)^3 2 P_H^0 \frac{d\sigma_H}{d^3P_H}= \sum_{n, n^\prime} \int \frac{d^4 P}{(2\pi)^4} d\hat{\sigma}_{[n n^\prime]}(P) F_{[n n^\prime]\to H}(P,P_H),
\end{align}
where $d\hat{\sigma}_{[n n^\prime]}(P)$ are perturbatively calculable short distance hard parts that produce a $Q \bar Q$ pair with quantum numbers $n=\state{{2S+1}}{L}{J,J_z}{c}$ and $n^\prime =\stateprime{{2S^\prime+1}}{L}{J^\prime,J_z^\prime}{c^\prime}$ in the amplitude and the complex-conjugate of the amplitude, respectively, with $c,c^\prime = 1 $ or $8$ representing the color-singlet or color-octet state of the $Q \bar Q$ pair. In general, $n$ can be different from $n^\prime$. In the
case of producing a polarization-summed quarkonium $H$, there are
constraints $c=c^\prime, S=S^\prime, J =J^\prime, J_z=J_z^\prime$ and $|L-L^\prime|=0,2,4,\cdots$~\cite{Ma:2017xno,Ma:2015yka}. In Eq.~\eqref{eq:fac4d}, $P$ is the total momentum of the intermediate $Q \bar Q$ pair, which is different from the momentum of the physical quarkonium $P_H$. The soft gluon distributions (SGDs), $F_{[n n^\prime]\to H}(P,P_H)$, are nonperturbative functions that describe the hadronization of the intermediate $Q\bar Q(n,n^\prime)$ states to heavy quarkonium $H$. 
The factorization formula Eq.~\eqref{eq:fac4d} is called 4-dimensional SGF (SGF-4d) in Ref.~\cite{Ma:2017xno}, and it can be simplified to the so called SGF-1d and SGF-0d formula by some further expansions. Using these SGF formulas, the authors studied the $J/\psi$ hadroproduction via gluon fragmenting into polarization-summed $\state{{3}}{S}{1}{8}$ intermediate state at tree level, and they found that the SGF-1d formula is a very good approximation of full SGF-4d formula.  The SGF approach has also been applied to exclusive quarkonium processes~\cite{Li:2019ncs,Chen:2020yeg}, in which the SGDs are reduced to local matrix elements.

In this paper we study quarkonium inclusive production using gluon fragmentation function (FF) $D_{g\rightarrow H}(z,\mu)$, where $z$ is the longitudinal momentum fraction of the quarkonium state and $\mu$ is the collinear factorization scale. To be concrete, we concentrate on the $\COcSa$  state production. In NRQCD framework, the SDC of gluon fragmenting to $\COcSa$ has been calculated up to next-to-leading order (NLO) in perturbative expansion~\cite{Ma:1995ci,Braaten:2000pc,Ma:2013yla} (but at the lowest order in velocity expansion).
It was found that SDC in NRQCD suffers from large (threshold) logarithms  of the form
\begin{equation}\label{eq:logarithm}
\alpha_s^i \biggr[\frac{\ln^j(1-z)}{1-z}\biggr]_+, \quad j\leq 2i,
\end{equation}
which spoils perturbative expansion in the region $z\to1$. Similar issues arise when analyzing the $J/\psi$ photoproduction and electroproduction, where the large logarithms were resummed to all orders in $\alpha_s$\cite{Fleming:2003gt,Fleming:2006cd,Bauer:2001rh,Beneke:1997qw} by combining the NRQCD with soft collinear effective theory (SCET)~\cite{Bauer:2000ew,Bauer:2000yr,Bauer:2001ct,Bauer:2001yt}. Although the same technique can also be used to resum the large logarithms appeared in FF, we choose to apply SGF instead of NRQCD+SCET to deal with this problem. The reason is that, as we will show, the SGF can not only resum the large logarithms but also resum a series of important relativistic corrections, which may result in better convergence for relativistic expansion.

The rest of the paper is organized as following. In Sec.~\ref{sec:SGF}, we review the SGF formula for quarkonium production, including the definition of SGDs, the perturbative matching procedure and the velocity expansion of short distance hard parts.
In Sec.~\ref{sec:SGDs}, we present the perturbative calculation of the SGDs.
In Sec.~\ref{sec:RGE}, we discuss the renormalization group equation (RGE) for the $\COcSa$ SGD.
In Sec.~\ref{sec:FF-SGF}, we study the gluon FF in SGF and calculate the short distance hard part in $\COcSa$ channel up to NLO. We conclude in Sec.~\ref{sec:Conclusion}. In appendix~\ref{sec:app-rge} we provide some details of solving the RGE of SGDs. In appendix~\ref{sec:app-angle}, we
provide some integral formulas used in our calculation. Finally, we list the expressions of finite functions that enter the gluon FF in appendix~\ref{sec:app-finite-function}.

\section{Soft gluon factorization}\label{sec:SGF}
Before studying the gluon fragmentation function, we first briefly review the
SGF formula for quarkonium production cross section. We denote $M_H$ and $m_Q$ as the mass of the heavy quarkonium $H$ and the mass of the heavy quark $Q$, respectively.
Lorentz-vector $a$, denoting as
 \begin{align}
  a^\mu= (a^0,a^1,a^2,a^3)=(a^0,\boldsymbol{a}),\nonumber
\end{align}
is sometimes also expressed in light-cone coordinates,
\begin{align}
  a^\mu&=(a^+,a^-,a^1,a^2)= (a^+,a^-,a_\perp),\nonumber\\
  a^+ &= (a^0+a^3)/\sqrt{2},\nonumber\\
  a^- &= (a^0-a^3)/\sqrt{2}.\nonumber
\end{align}
The scalar product of two four-vector $a$ and $b$ then becomes
\begin{align}
  a \cdot b = a^+ b^- + a^- b^+ + a_\perp \cdot b_\perp. \nonumber
\end{align}
We introduce a light-like vector $l^\mu = (0, 1, 0_\perp)$ so that $a\cdot l=a^+$.

Following the discussion in~\cite{Ma:2017xno}, the factorization formula Eq.~\eqref{eq:fac4d} for producing a quarkonium $H$ with momentum $P_H$ can be simplified to the following SGF-1d formula
\begin{align}\label{eq:fac1d-1}
 (2\pi)^3 2 P_H^0 \frac{d\sigma_H}{d^3P_H}
 &= \sum_{n,n^\prime} \int \frac{dz}{z^2}   d\hat{\sigma}_{[n n^\prime ]}(P_H/z, m_Q, \mu_f)    F_{[n n^\prime] \rightarrow H }(z,M_H,m_Q, \mu_f),
\end{align}
which has similar good convergence in velocity expansion for our purpose, and at the same time is more suitable in practical use. In the above formula, $d\hat{\sigma}_{[n n^\prime]}(P_H/z, m_Q, \mu_f)$ are the short-distance hard parts which, roughly speaking, produce a $Q\bar{Q}$ pair with momentum $P_H/z$ and quantum number $n$ in the amplitude and $n^\prime$ in the complex conjugate of the amplitude, $F_{[n n^\prime] \to H }(z,M_H,m_Q, \mu_f)$ are the one-dimensional SGDs which describe the hadronization from the $Q\bar{Q}$ pair to quarkonium $H$,  $\mu_f$ is the factorization scale and $z = P^+_H/P^+$ is the longitudinal momentum fraction with $P$ denoting the total momentum of the intermediate $Q\bar Q$ pair.

SGDs are defined as
\begin{align}\label{eq:SGD1d-1}
F_{[n n^\prime] \to H}(z,M_H,m_Q, \mu_f)
&= P_H^+\int \frac{db^-}{2\pi} e^{-iP_H^+  b^-/z} \langle 0| [\bar\Psi \mathcal {K}_{n} \Psi]^\dag(0) [a_H^\dag a_H] [\bar\Psi \mathcal {K}_{n^\prime}\Psi](b^-) |0\rangle_{\textrm{S}},
\end{align}
where $\Psi$	stands for Dirac field of heavy quark and the subscript ``S'' means the field operators in the above definition are the operators obtained in small momentum region. In additional, in this paper we define ``S'' to select only leading power terms in threshold $(P-P_H)^+=(1-z)P^+$ expansion, which is sufficient to factorize and then resum leading-power large logarithms as $z\to 1$. 
In Eq.~\eqref{eq:SGD1d-1}, the quarkonium state, created by $a_H^\dagger$, has standard relativistic normalization and the projection operators $\mathcal {K}_{n}$ can be decomposed to a spin operator, a color operator, and a gauge link~\cite{Ma:2017xno}. For the case $n=\state{{3}}{S}{1,\lambda}{8}, \state{{3}}{D}{1,\lambda}{8}$ which will be studied in this work, we have
\begin{subequations}\label{eq:projection1}
\begin{align}
 \mathcal {K}_{\state{{3}}{S}{1,\lambda}{8}}(rb^-) =& \frac{\sqrt{M_ {H}}}{M_ {H}+2 m}\frac{M_ {H} + \slashed{P}_ {H}}{2M_ {H}} \mathcal {C}^{[8]}_{ a} {\epsilon}_{\lambda}^{\mu}\gamma_\mu \frac{M_ {H} - \slashed{P}_ {H}}{2M_ {H}} ,
 \\
 \mathcal {K}_{\state{{3}}{D}{1,\lambda}{8}}(rb^-) =& \frac{\sqrt{M_ {H}}}{M_ {H}+2 m}\frac{M_ {H} + \slashed{P}_ {H}}{2M_ {H}} \mathcal {C}^{[8]}_{ a} {\epsilon}_{\lambda}^{\mu}\gamma^\nu \biggr(-\frac{i}{2}\biggr)^2 \lrd^\alpha\lrd^\beta \frac{M_ {H} - \slashed{P}_ {H}}{2M_ {H}}
 \nonumber\\
 &\times  \sqrt{\frac{2(d-1)}{(d-2)(d+1)}} \biggr( - \mathbb{P}_{\alpha\mu}\mathbb{P}_{\beta\nu} + \frac{1}{d-1} \mathbb{P}_{\alpha\beta} \mathbb{P}_{\mu\nu} \biggr),
\end{align}
\end{subequations}
where $D_\mu$ is the gauge covariant derivative with $\overline\Psi \lrd_\mu \Psi =
\overline\Psi (D_\mu \Psi) -
(D_\mu \overline\Psi)\Psi$ and  $\epsilon_{\lambda}$ are polarization vectors which satisfy the following relations
\begin{align}
P_H \cdot \epsilon_{\lambda} =0, \qquad
\epsilon_{\lambda}\cdot \epsilon_{\lambda^\prime}^\ast =-\delta_{\lambda \lambda^\prime}, \qquad
\sum_{\lambda} \epsilon_{\lambda}^\mu \epsilon_{\lambda}^{\ast\nu} = \mathbb{P}_{\mu\nu}.
\end{align}
The spin projection operator $\mathbb{P}_{\alpha\beta}$ is defined as
\begin{align}
\mathbb{P}_{\alpha\beta}=-g_{\alpha\beta}+\frac{P_{H\alpha} P_{H\beta}}{M_H^2}.
\end{align}
The color operator is defined as
\begin{align}
\mathcal {C}^{[8]}_{ a}=\sqrt{2}T^{ \bar a} \Phi_{l}(rb^-)_{\bar a  a}.
\end{align}
 The gauge link $\Phi_{l}(rb^-)_{\bar a   a}$ is introduced to enable gauge invariance of SGD, which is defined along the $l^\mu$ direction,
\begin{equation}\label{eq:gaugelink}
  \Phi_l(rb^{-})= \mathcal {P} \, \text{exp} \left[-i g_s
  \int_{0}^{\infty}\mathrm{d}\xi l\cdot A(rb^{-} + \xi l) \right] \, ,
\end{equation}
where $\mathcal {P}$ denotes path ordering and $A^{\mu}(x)$ is the gluon field in the adjoint representation: $[A^{\mu}(x)]_{ac} = i f^{abc} A^{\mu}_{b}(x)$.

As the short distance hard parts $d\hat{\sigma}_{nn^\prime}(P_H/z, m_Q, \mu_f)$ do not depend on nonperturbative
physics, they can be  perturbatively calculated. To this end, we replace the quarkonium $H$ in Eq.~\eqref{eq:fac1d-1} by an on-shell $Q \bar Q$ pair with certain quantum number $m$ in the amplitude and $m^\prime$ in the complex-conjugate amplitude, which results in
\begin{align}\label{eq:matchingEq1}
d \bar{\sigma}_{Q\bar Q[m m^\prime]}&=  \sum_{n,n^\prime} \int \frac{dz}{z^2}  d\hat{\sigma}_{[n n^\prime]}(P_H/z, m_Q, \mu_f)  F_{[n n^\prime] \to Q\bar Q[m m^\prime] }(z,M_H,m_Q, \mu_f),
\end{align}
where $d \bar{\sigma}_{Q\bar Q[m m^\prime]} \equiv (2\pi)^3 2 P_H^0 d\sigma_{Q\bar Q[m m^\prime]} / d^3P_H$.
We can determine the hard parts by computing both sides of Eq.~\eqref{eq:matchingEq1} in perturbation theory. If the factorization Eq.~\eqref{eq:fac1d-1} holds, the obtained hard parts will be free of infrared divergences. Momenta of the on-shell $Q \bar Q$ pair are chosen as
\begin{align}
p_Q= \frac{1}{2}P_H +q ,  \quad \quad p_{\bar Q}= \frac{1}{2}P_H -q.
\end{align}
On-shell conditions $p_Q^2= p_{\bar Q}^2=m_Q^2$ result in
\begin{align}
P_H\cdot q= 0 , \quad \quad q^2=m_Q^2-P_H^2/4.
\end{align}
To project the final state $Q \bar Q$ pair to quantum number $m(\state{{2S +1}}{L}{J,J_z }{c })$, we replace the spinors of $Q \bar Q$ by the following projector
\begin{align}\label{eq:spinor-operator}
\Pi[m]=\frac{2}{\sqrt{M_H}(M_H+2m_Q)} ( \slashed{p}_{\bar Q} - m_Q )  \frac{M_H - \slashed{P}_H}{2M_H} \widetilde{\Gamma}_{m }
\frac{M_H + \slashed{P}_H}{2M_H} (\slashed{p}_{Q} +m_Q ).
\end{align}
For $m=\state{{3}}{S}{1,\lambda}{8}$ or $\state{{3}}{D}{1,\lambda}{8}$, we have
\begin{align}
\widetilde{\Gamma}_{m } =  & \widetilde{\Gamma}^s_{m}
\widetilde{\mathcal {C}}^{[8],a}.
\end{align}
The color operators and spin operator have similar definitions as those in Eq.~\eqref{eq:projection1}, which are given by
\begin{subequations}
\begin{align}
\widetilde{\mathcal {C}}^{[8],a}=&\sqrt{\frac{2}{N_c^2-1}}T^{ a}, \\
\widetilde{\Gamma}^s_{\state{{3}}{S}{1,\lambda}{8}}=&\epsilon_{\lambda}^{\ast\mu} \gamma_\mu, \\
\widetilde{\Gamma}^s_{\state{{3}}{D}{1,\lambda}{8}}=&  \frac{(d-1)(d+1)}{2 } \frac{q^\mu q^\nu}{\vert \textbf{q} \vert^4} \epsilon_{\lambda}^{\ast \beta} \gamma^\alpha
\sqrt{\frac{d-1}{2(d-2)(d+1)}} \nonumber\\
 & \times \biggr( -\mathbb{P}_{\alpha\mu}\mathbb{P}_{\beta\nu} -\mathbb{P}_{\alpha\nu}\mathbb{P}_{\beta\mu} + \frac{2}{d-1} \mathbb{P}_{\alpha\beta} \mathbb{P}_{\mu\nu} \biggr),
\end{align}
\end{subequations}
where the factor $\sqrt{N_c^2-1}$ is to average over
color-octet states.

According to the definition in Eq.~\eqref{eq:SGD1d-1}, with $H$ replaced by $Q\bar{Q}[m m^\prime]$, and the projector in Eq.\eqref{eq:spinor-operator}, it can be found that, at the lowest order in $\alpha_s$, the free $Q \bar Q$ SGDs have the orthogonality relations~\cite{Ma:2017xno}
\begin{align}\label{eq:SGD-LO}
F^{LO}_{[nn^\prime] \rightarrow Q\bar{Q}[m m^\prime] }(z,M_H,m_Q, \mu_f) &= \delta_{nm}\delta_{n^\prime m^\prime}\delta(1-z).
\end{align}
By inserting perturbative expansions
\begin{align}
F_{[n n^\prime] \to Q\bar Q[m m^\prime]} =& F_{[n n^\prime] \to Q\bar Q[m m^\prime]}^{LO} + F_{[n n^\prime] \to Q\bar Q[m m^\prime]}^{NLO} + \cdots,
\nonumber\\
d \bar{\sigma}_{Q\bar Q[m m^\prime]} =& d \bar{\sigma}_{Q\bar Q[m m^\prime]}^{LO} + d \bar{\sigma}_{Q\bar Q[m m^\prime]}^{NLO} + \cdots,
\nonumber\\
d \hat{\sigma}_{[n n^\prime]} =& d \hat{\sigma}^{LO}_{[n n^\prime]} +  d \hat{\sigma}^{NLO}_{[n n^\prime]} + \cdots,
\end{align}
into Eq.~\eqref{eq:matchingEq1} and using the orthogonal relations
Eq.~\eqref{eq:SGD-LO}, one obtains relations
\begin{align}\label{eq:relation}
d\hat{\sigma}_{[n n^\prime]}^{LO}(P_H/z, m_Q, \mu_f) =& d \bar{\sigma}_{Q\bar Q[n n^\prime]}^{LO}(P_H/z, m_Q),
\nonumber\\
d\hat{\sigma}_{[n n^\prime]}^{NLO}(P_H/z, m_Q, \mu_f) =&  d \bar{\sigma}_{Q\bar Q[n n^\prime]}^{NLO}(P_H/z,m_Q)
\nonumber\\
&  \hspace{-4cm}
- \sum_{m,m^\prime} \int \frac{dx}{x^2}  d\hat{\sigma}_{[m m^\prime]}^{LO}(P_H/(xz),m_Q, \mu_f)  F_{[mm^\prime] \to Q\bar Q[n n^\prime] }^{NLO}(x,M_H/z,m_Q, \mu_f),
\end{align}
and so on, which express short-distance hard parts in terms of perturbative calculated $d \bar{\sigma}_{Q\bar Q[n n^\prime]}$ and $F_{[n n^\prime] \to Q\bar Q[m m^\prime] }$.

Finally, we note that perturbative calculation with analytical $q$ dependence are not very easy for complicated processes. There are at least two independent hard scales, $m_Q$ and $M_H/z$ in the hard parts. As suggested in~\cite{Ma:2017xno}, we can further simplify the hard parts by expanding $m_Q^2$ around $M_H^2/(4z^2)$, i.e.
\begin{align}
d\hat{\sigma}_{[n n^\prime]}(P_H/z,m_Q, \mu_f)&= d\hat{\sigma}_{[n n^\prime]}(P_H/z, M_H/(2z),  \mu_f)
 \nonumber\\
&  \hspace{-2.5cm} ~~~+ \frac{\partial [d\hat{\sigma}_{[n n^\prime]}(P_H/z,m_Q, \mu_f)]}{\partial m_Q^2} \biggr \vert_{m_Q^2 =M_H^2/(4z^2)} \biggr( m_Q^2 -\frac{M_H^2}{4z^2} \biggr) + \cdots \nonumber\\
& \hspace{-2.5cm} =  \sum_{i=0} d\hat{\sigma}_{[n n^\prime]}^{(i)}(P_H/z, \mu_f)  \tilde {q}^{2i},
\end{align}
where $\tilde {q}^{2}=m_Q^2 -M_H^2/(4z^2)\equiv m_Q^2 v^2$ with $v$ being the velocity of the heavy quark in the rest frame of the pair. Then the SGF formula Eq.~\eqref{eq:fac1d-1} can be rewritten as
\begin{align}\label{eq:fac1d-2}
 d \bar{\sigma}_{H}&=  \sum_{n n^\prime} \int \frac{dz}{z^2}  \sum_{i=0} d\hat{\sigma}_{[n n^\prime]}^{(i)}(P_H/z, \mu_f)  \tilde {q}^{2i}  F_{[n n^\prime] \to H }(z,M_H,m_Q, \mu_f),
\end{align}
which defines a velocity expansion series in SGF. The tree level calculation in Ref.~\cite{Ma:2017xno} shows that the
convergence of velocity expansion in SGF is much better than that in NRQCD. We will confirm this conclusion at one-loop level in this paper.

\section{Perturbative calculation of soft gluon distributions}\label{sec:SGDs}

\subsection{Definition}

For the purpose of this paper, we are interested in the intermediate state $n$ or $n^\prime$ equals to state $\state{{3}}{S}{1,\lambda}{8}$ or $\state{{3}}{D}{1,\lambda}{8}$ (denoted as $S$ and $D$ respectively). For convenient, we denote
\begin{subequations}
	\begin{align}
	[L\tilde{L},\lambda]& \equiv[\state{{3}}{L}{1,\lambda}{8}\state{{3}}{\tilde{L}}{1,\lambda}{8}],\\
	[L\tilde{L}]&\equiv[\state{{3}}{L}{1}{8}\state{{3}}{\tilde{L}}{1}{8}].
	\end{align}
\end{subequations}
The notations in the second line denote polarization-summed intermediate states.
In general, the SGD $F_{[L\tilde{L},\lambda] \to H}$ is $\lambda$ dependent even for polarization-summed $H$. This is different to the polarized NRQCD LDMEs, which can be simplified to the usual unpolarized NRQCD LDMEs due to the rotation invariance~\cite{Ma:2015yka,Braaten:1996jt}. But in the case of SGD, the $z$-axis direction needs to be specified because the longitudinal momentum of intermediate $Q \bar Q$ pair is fixed, thus the rotation invariance is broken.
Similar to the polarized NRQCD
LDMEs defined in~\cite{Ma:2015yka}, it is convenient to construct the definitions of polarized SGDs as
\begin{subequations}
	\begin{align}\label{eq:pol-SGD}
	F_{[L\tilde{L},T] \to H}(z,M_H,m_Q, \mu_f) &= \frac{1}{d-2}\sum_{\vert \lambda \vert =1} F_{[L\tilde{L},\lambda] \to H}(z,M_H,m_Q, \mu_f),  \\
	F_{[L\tilde{L},L] \to H}(z,M_H,m_Q, \mu_f) &=  \sum_{\vert \lambda \vert =0} F_{[L\tilde{L},\lambda] \to H}(z,M_H,m_Q, \mu_f).
	\end{align}
\end{subequations}
But for the polarization-summed $S$-wave quarkonium $H$, such as $J/\psi$, the polarized $\COcSa$ SGDs can be reduced to following unpolarized SGD:
\begin{align}
F_{[SS] \to H}(z,M_H,m_Q, \mu_f) &= \frac{1}{d-1}\sum_{\lambda} F_{[SS,\lambda] \to H}(z,M_H,m_Q, \mu_f).
\end{align}
And we have
\begin{align}
\begin{split}\label{eq:reduce}
F_{[SS,T] \to H}(z,M_H,m_Q, \mu_f) &= F_{[SS,L] \to H}(z,M_H,m_Q, \mu_f) \\
&= F_{[SS] \to H}(z,M_H,m_Q, \mu_f).
\end{split}
\end{align}
We will explain this later.

After replacing $H$ with $Q\bar{Q}[L\tilde{L},\lambda]$, the general form of $F_{[L^\prime \tilde{L}^\prime, \lambda^\prime] \to Q\bar Q[L\tilde{L},\lambda]}$ can be written as
\begin{align}\label{eq:genSGD}
F_{ [L^\prime \tilde{L}^\prime, \lambda^\prime] \to Q\bar Q[L\tilde{L},\lambda] }
(z,M_H,m_Q,\mu_f) =& T_S\biggr[P_H^+ \int
d \Phi \,   \mathcal{A}_{L^\prime L}(P_H
,P,m_Q,k_i,\lambda^\prime,\lambda)
\nonumber \\
&\times
\mathcal{A}_{\tilde{L}^\prime \tilde{L}}^*(P_H
,P,m_Q,k_i,\lambda^\prime,\lambda) \biggr]  ,
\end{align}
where $d \Phi$ is the final-state phase space, $k_i$ denotes momenta of the
final-state gluons or light (anti-)quarks, and $\mathcal{A}_{L^\prime L}(P_H
,P,m_Q,k_i,\lambda^\prime,\lambda)$ denotes the amplitude of transition from a $Q \bar Q$ pair in $\stateprime{{3}}{L}{1,\lambda^\prime}{8}$ state into a $Q\bar Q$ pair in $\state{{3}}{L}{1,\lambda}{8}$ state which can be
written as
\begin{align}\label{eq:SGFAM}
\mathcal{A}_{L^\prime L}(P_H
,P,m_Q,k_i,\lambda^\prime,\lambda) =& \int \frac{d^{d-2}\Omega }{N_\Omega}     \textrm{Tr} \biggr[ \mathcal{A}_{L^\prime}^{Q\bar Q} \Pi[\state{{3}}{L}{1,\lambda}{8}]   \biggr] ,
\end{align}
where $\mathcal{A}_{L^\prime}^{Q\bar Q}$ is the amplitude to produce an open $Q\bar{Q}$ pair with spinors of the pair removed. As pointed out in the last section, in addition to force loop momenta and light-parton momenta $k_i$ in small-momentum region, the operator $T_S$ also selects only the leading power contributions in the threshold expansion.

Based on Eq.~\eqref{eq:genSGD}, we find that, for polarization-summed final state, SGDs can be decomposed as
\begin{align}
\sum_{\lambda} F_{ [L^\prime \tilde{L}^\prime, \lambda^\prime] \to Q\bar Q[L\tilde{L},\lambda] }
(z,M_H,m_Q,\mu_f) =&  \biggr(F_1 \mathbb{P}_{\mu\nu}+ F_2 l_{\mu}l_{\nu} +F_3(P_{H\mu}l_{\nu}+P_{H\nu}l_{\nu})
\nonumber\\
&+F_4 P_{H\mu}P_{H\nu}\biggr){\epsilon}_{\lambda^\prime}^{\mu}{\epsilon}_{\lambda^\prime}^{\nu}.
\end{align}
For general SGDs, $F_2$ and $F_3$ are nonzero and the existence of $l^\mu$ breaks the rotation invariance. But for the $S$-wave case, $F_2$ and $F_3$ are zero thanks to keeping only LP term in threshold expansion, which results in
\begin{align}
\begin{split}
F_{[SS,T] \to Q\bar Q[SS]}(z,M_H,m_Q, \mu_f) &= F_{[SS,L] \to Q\bar Q[SS]}(z,M_H,m_Q, \mu_f) \\
&= F_{[SS] \to Q\bar Q[SS]}(z,M_H,m_Q, \mu_f).
\end{split}
\end{align}
Thus we argue that for the polarization-summed $S$-wave quarkonium $H$, we have the relations in Eq.~\eqref{eq:reduce}.

\subsection{LO calculation}

\begin{figure}[htb!]
	\begin{center}
		\includegraphics[width=0.4\textwidth]{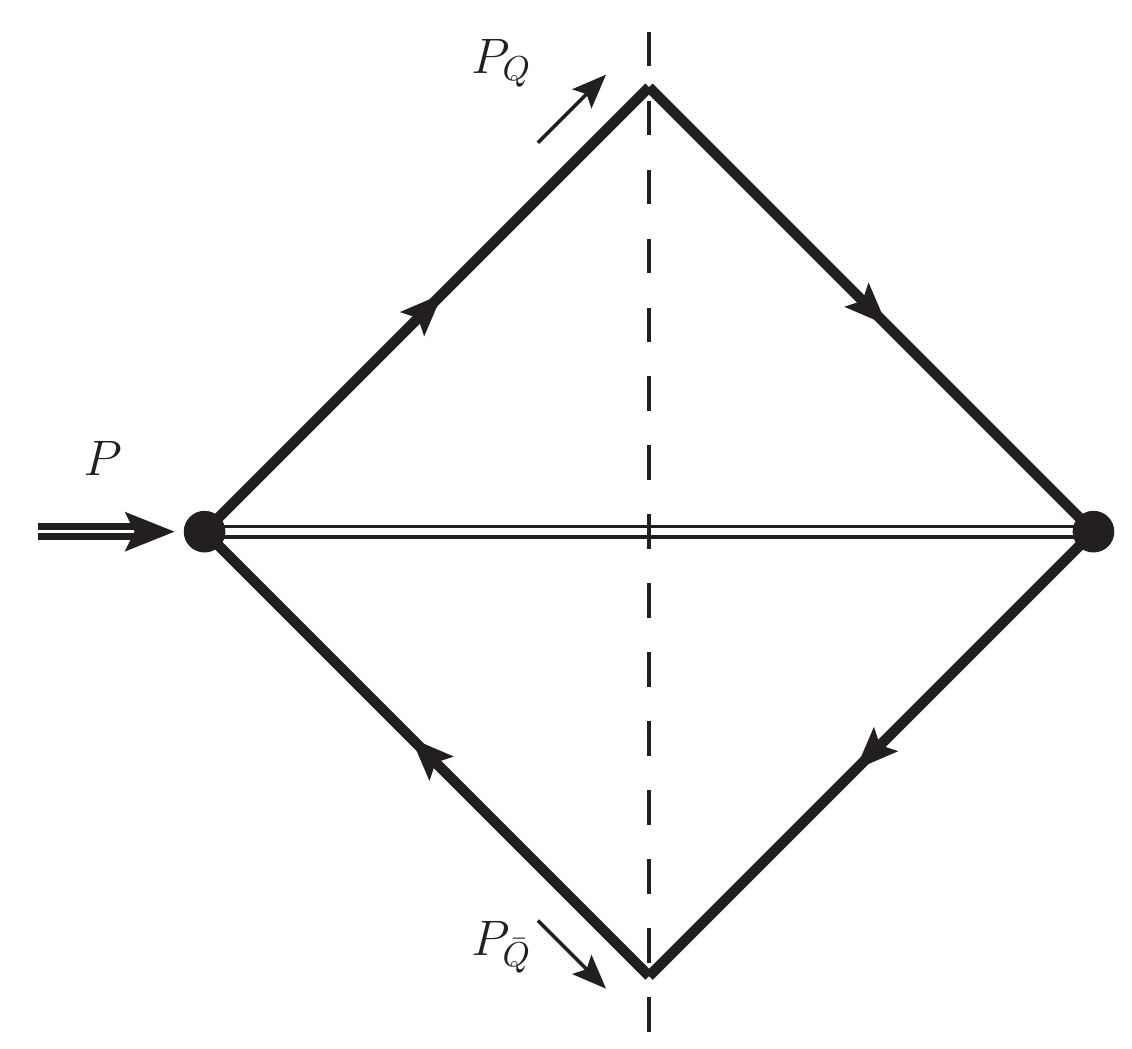}
		\caption{Feynman diagrams for the free $Q \bar Q$ SGD at LO. The double solid line represents the gauge link along $l$ direction. \label{fig:LOSGD}}
	\end{center}
\end{figure}

The LO Feynman diagram for $F_{[SS] \to Q\bar Q[SS]}$ is shown in Fig.~\ref{fig:LOSGD}, where the solid circles represent the operator $\overline\Psi {\cal K}_{\state{{3}}{S}{1,\lambda}{8}}\Psi$.
Based on Eqs.~\eqref{eq:genSGD}, \eqref{eq:projection1} and \eqref{eq:spinor-operator} we obtain
\begin{align}\label{eq:SGD-LO0}
F_{[SS] \to Q\bar Q[SS]}^{LO}(x,M_H,m_Q,\mu_f)= & \frac{1}{d-1}\sum_{\lambda^\prime,\lambda} F_{[SS,\lambda^\prime] \to Q\bar Q[SS,\lambda]}^{LO}(x,M_H,m_Q,\mu_f)\nonumber\\
&\hspace{-4cm} =\frac{P_H^+}{d-1} \int \frac{d^d P}{(2\pi)^d} \delta(P^+ - \frac{P_H^+}{x}) (2\pi)^d \delta^{d}(P - P_H)  \mathcal {M}_S^{\alpha\beta} \mathcal {M}_S^{\ast\rho \sigma}  \mathbb{P}_{\alpha\rho} \mathbb{P}_{\beta \sigma},
\end{align}
with
\begin{align}
\mathcal {M}^{\alpha\beta}_S =&  \int \frac{d^{d-2}\Omega}{N_\Omega}
\textrm{Tr} [  \mathcal {C}_8^{\prime b} \Pi_3^{\prime \alpha}
\widetilde{\mathcal {C}}^{[8],d} \Pi_3^{\beta} ],
\end{align}
where
\begin{align}
\mathcal {C}_8^{\prime b}=& \sqrt{2}T^b,
\nonumber\\
\Pi_3^{\prime \alpha} =&  \frac{\sqrt{M_H}}{M_H+2m_Q}   \frac{M_H + \slashed{P}_H}{2M_H} \gamma^\alpha
\frac{M_H - \slashed{P}_H}{2M_H},
\nonumber\\
\Pi_3^\beta=& \frac{2}{\sqrt{M_H}(M_H+2m_Q)} ( \slashed{p}_{\bar Q} - m_Q )  \frac{M_H - \slashed{P}_H}{2M_H} \gamma^\beta
\frac{M_H + \slashed{P}_H}{2M_H} (\slashed{p}_{Q} +m_Q ).
\end{align}
Then we obtain
\begin{align}\label{eq:SGD-LO1}
F_{[SS] \to Q\bar Q[SS]}^{LO}(x,M_H,m_Q,\mu_f)  =& \delta(1-x),
\end{align}
which is consistent with the result in Eq.~\eqref{eq:SGD-LO}. Similarly we can derive
\begin{align}\label{eq:SGD-LO2}
F_{[SD,T] \to Q\bar Q[SD,T]}^{LO}(x,M_H,m_Q,\mu_f)  =& \delta(1-x), \nonumber\\
F_{[DD,T] \to Q\bar Q[DD,T]}^{LO}(x,M_H,m_Q,\mu_f)  =& \delta(1-x).
\end{align}
To obtain the above results, we have used the relations
\begin{subequations}
\begin{align}
\sum_{
|\lambda|=1} \epsilon_{\lambda}^\mu \epsilon_{\lambda}^{\ast\nu} =&
-g^{\mu\nu}+\frac{P_H^\mu l^\nu +l^\mu P_H^\nu}{P_H\cdot l} - \frac{M_H^2 l^\mu l^\nu  }
{(P_H\cdot l)^2}, \\
\sum_{
|\lambda|=0} \epsilon_{\lambda}^\mu \epsilon_{\lambda}^{\ast\nu} =&
\frac{ P_H^\mu P_H^\nu  }{M_H^2} -\frac{P_H^\mu l^\nu +l^\mu P_H^\nu}{P_H\cdot l}
+ \frac{M_H^2 l^\mu l^\nu  }{(P_H\cdot l)^2}.
\end{align}
\end{subequations}
The longitudinally polarized ($\lambda=L$) SGDs can be calculated similarly, but they are irrelevant for our purpose as we will show later, thus we do not consider their contributions in our calculation.

\subsection{NLO calculation}

\begin{figure}[htb!]
	\begin{center}
		\includegraphics[width=0.8\textwidth]{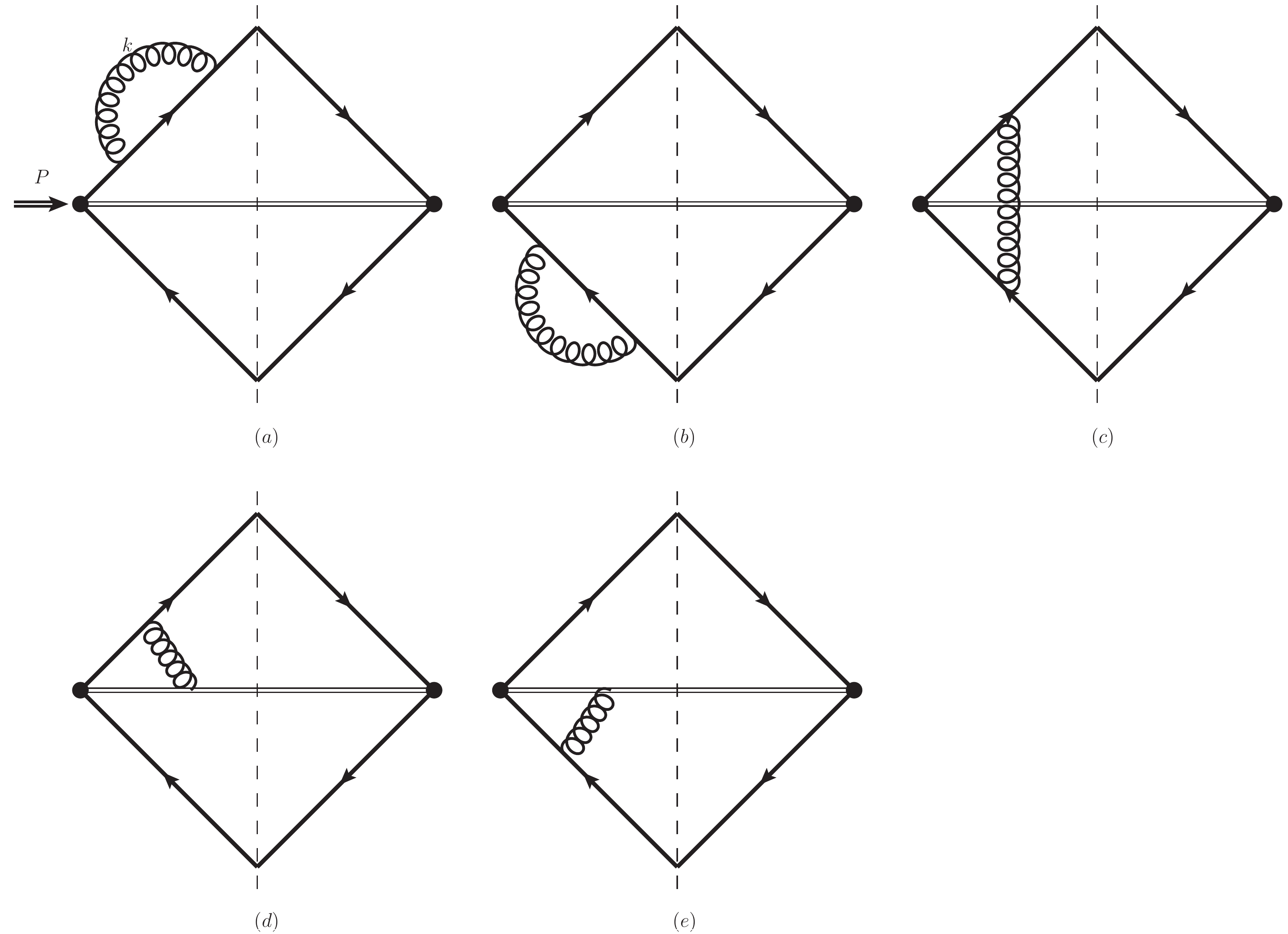}
		\caption{The virtual corrections of SGD at NLO. The complex conjugate diagrams are
			not shown, but are included in the calculations. \label{fig:SGD-virtual}}
	\end{center}
\end{figure}

\begin{figure}[htb!]
	\begin{center}
		\vspace*{0.8cm}
		\hspace*{-5mm}
		\includegraphics[width=0.8\textwidth]{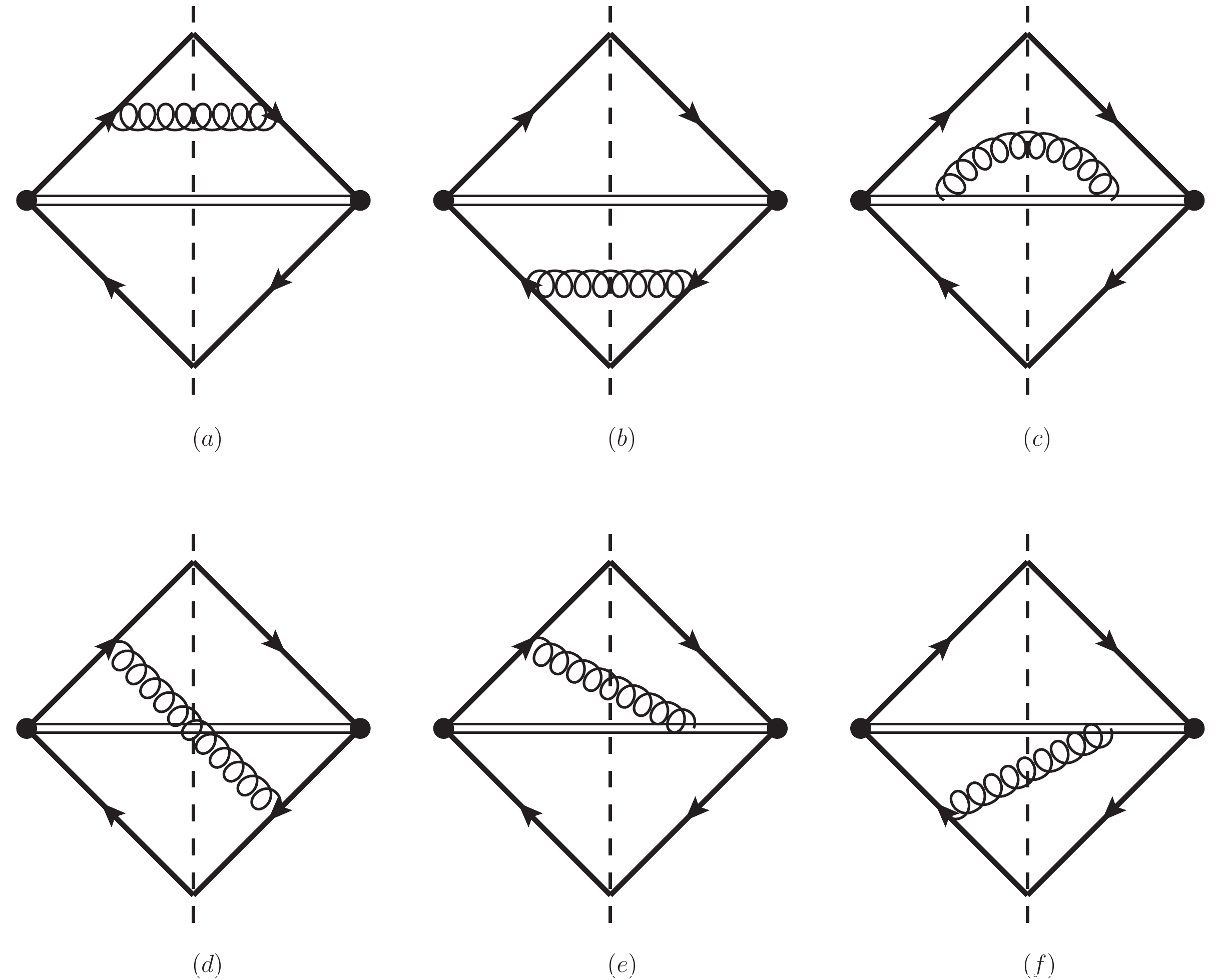}
	\end{center}
	\vspace*{-.5cm}
	\caption{The real corrections of SGD at NLO.}\label{fig:SGD-real}
	\vspace*{0.cm}
\end{figure}

At NLO, the corresponding Feynman diagrams are shown in
Fig.~\ref{fig:SGD-virtual} and Fig.~\ref{fig:SGD-real}.
Let us first consider virtual corrections in Fig.~\ref{fig:SGD-virtual}. Following the calculation in ref.~\cite{Chen:2020yeg}, we obtain
\begin{subequations}\label{eq:SGD-vir}
	\begin{align}
	F_{[SS] \to Q\bar Q[SS]}^{NLO}(x,M_H,m_Q,\mu_f)\biggr \vert^{a+b}_{vir.} =&
	\frac{\alpha_s}{4\pi} \frac{N_c^2-1}{2N_c} \biggr(\frac{4}{\epsilon_{UV}} - \frac{4}{\epsilon_{IR}} \biggr)
	\delta(1-x),\\
	F_{[SS] \to Q\bar Q[SS]}^{NLO}(x,M_H,m_Q,\mu_f) \biggr \vert^{c}_{vir.} =&
	\frac{\alpha_s}{4\pi} \frac{1}{2N_c}  \biggr[\biggr(\frac{4}{\epsilon_{UV}} - \frac{4}{\epsilon_{IR}} \biggr)
	\frac{1+\Delta^2}{2\Delta} \ln \biggr( \frac{1+\Delta}{1-\Delta} \biggr)
	\nonumber\\
	& - 2(1+\Delta^2)\frac{\pi^2}{\Delta} \biggr]\delta(1-x) ,\\
	F_{[SS] \to Q\bar Q[SS]}^{NLO}(x,M_H,m_Q,\mu_f) \biggr \vert^{d+e}_{vir.} =&
	\frac{\alpha_s}{4\pi} \frac{N_c}{2}(-1)\biggr( \frac{\pi\mu^2 e^{-\gamma_E}}{E^2} \biggr)^{\epsilon}\biggr[
	\frac{1}{\epsilon_{UV}} \biggr(\frac{4}{\epsilon_{UV}} - \frac{4}{\epsilon_{IR}} \biggr)   \nonumber\\
	& + 2\biggr( \frac{1}{2\Delta}\ln\frac{1+\Delta}{1-\Delta} - 1 \biggr)\biggr(\frac{4}{\epsilon_{UV}} -
	\frac{4}{\epsilon_{IR}} \biggr) \biggr]
	\nonumber\\
	& \times \delta(1-x) .
	\end{align}
\end{subequations}
with
\begin{equation}
	\Delta =\frac{\sqrt{M_H^2-4m_Q^2}}{M_H}.
\end{equation}

For real corrections in Fig.\ref{fig:SGD-real}, we have
\begin{align}\label{eq:general-SGD-real}
F^{(NLO)}_{[SS] \to Q\bar Q[SS]}(z,M_H,m_Q,\mu_f)\biggr|_{real}
&= T_S \biggr[ \int \frac{d^{d}k}{(2\pi)^d} I^\prime \biggr],
\end{align}
with
\begin{align}
\begin{split}
I^\prime =&\frac{P_H^+}{d-1}
\delta (k^+ - w )
2\pi \delta(k^2) \theta(k^+)
\mathcal {M}^{\alpha\beta\sigma}_{S,R} \mathcal {M}^{\ast\mu \nu \rho}_{S,R}
\mathbb{P}_{\alpha\mu} \mathbb{P}_{\beta \nu}(-g_{\rho\sigma}), \quad w=\frac{1-x}{x}P_H^+,
\end{split}
\end{align}
where
\begin{align}
\mathcal {M}^{\alpha\beta\sigma}_{S,R} =& \int \frac{d^{d-2}\Omega}{N_\Omega}
\bigg( \mathcal {M}^{\alpha\beta\sigma,(1)}_{S,R} +\mathcal {M}^{\alpha\beta\sigma,(2)}_{S,R}
+\mathcal {M}^{\alpha\beta\sigma,(3)}_{S,R} \bigg) , \nonumber\\
\mathcal {M}^{\alpha\beta\sigma,(1)}_{S,R} =& g_s\mu^{\epsilon}\frac{\textrm{Tr}
	[\gamma^\sigma T^a  (\slashed{p}_{{ Q }}+\slashed{k}+m_Q) \mathcal {C}_8^{\prime b}
	\Pi_3^{\prime \alpha}
	\widetilde{\mathcal {C}}^{[8],d} \Pi_3^{\beta} ]}{2p_{ Q}\cdot k +i\varepsilon},
\nonumber\\
\mathcal {M}^{\alpha\beta\sigma,(2)}_{S,R} =& g_s\mu^{\epsilon}\frac{\textrm{Tr}
	[\mathcal {C}_8^{\prime b}
	\Pi_3^{\prime \alpha} (-\slashed{p}_{{ \bar Q }}-\slashed{k}+m) \gamma^\sigma T^a
	\widetilde{\mathcal {C}}^{[8],d} \Pi_3^{\beta} ]}{2p_{ \bar Q}\cdot k +i\varepsilon},
\nonumber\\
\mathcal {M}^{\alpha\beta\sigma,(3)}_{S,R} =& g_s \mu^{\epsilon} \textrm{Tr}
[\mathcal {C}_8^{\prime b}
\Pi_3^{\prime \alpha}
\widetilde{\mathcal {C}}^{[8],d} \Pi_3^{\beta} ] \biggr(i f^{abc}
\frac{l^\sigma}{l \cdot k + i\varepsilon}\biggr).
\end{align}
Here $k$ is the momentum of the emitted real gluon in the final state, which is restricted in the soft domain. Applying $T_S$ to $I^\prime$, we obtain
\begin{align}
T_S I^\prime =& \frac{P_H^+}{d-1}
\delta (k^+ - w )
2\pi \delta(k^2) \theta(k^+) \mathbb{P}_{\alpha\mu} \mathbb{P}_{\beta \nu}(-g_{\rho\sigma})
T_S \biggr\{\mathcal {M}^{\alpha\beta\sigma}_{S,R} \mathcal {M}^{\ast\mu \nu \rho}_{S,R}  \biggr\},
\end{align}
with
\begin{align}
T_S \mathcal {M}^{\alpha\beta\sigma,(1)}_R =& g_s\mu^{\epsilon} \textrm{Tr} [ T^a \mathcal {C}_8^{\prime b} \Pi_3^{\prime \alpha}
\widetilde{\mathcal {C}}^{[8],d} \Pi_3^{\beta} ] \sum_{j} \frac{(p^\sigma + 2q^\sigma) (2\boldsymbol{q} \cdot \boldsymbol{k})^j }{[p_0k_0 +i\varepsilon]^{1+j}}, \nonumber\\
T_S \mathcal {M}^{\alpha\beta\sigma,(2)}_R=&  g_s\mu^{\epsilon} \textrm{Tr} [  \mathcal {C}_8^{\prime b} \Pi_3^{\prime \alpha} T^a
\widetilde{\mathcal {C}}^{[8],d} \Pi_3^{\beta} ] \sum_{j} \frac{-( p^\sigma - 2q^\sigma) (-2\boldsymbol{q} \cdot \boldsymbol{k})^j }{[p_0k_0 +i\varepsilon]^{1+j}}, \nonumber\\
T_S \mathcal {M}^{\alpha\beta\sigma,(3)}_R=& g_s \mu^{\epsilon} \textrm{Tr} [  \mathcal {C}_8^{\prime b} \Pi_3^{\prime \alpha}
\widetilde{\mathcal {C}}^{[8],d} \Pi_3^{\beta} ] \biggr(i f^{abc}\frac{l^\sigma}{l \cdot k + i\varepsilon}\biggr).
\end{align}
Performing the $\Omega$ and $k$ integration, we obtain the following result
\begin{align}\label{eq:SGD-real}
F^{NLO}_{[SS] \to Q\bar Q[SS]}(x,M_H,m_Q,\mu_f) \biggr\vert_{real} =&
\frac{\alpha_s N_c}{4\pi}\biggr (\frac{4\pi x^2  \mu^2 e^{-\gamma_E}}{ M_H^2} \biggr)^{\epsilon} 4x(1-x)^{-1-2\epsilon} \nonumber\\
& \times
\biggr(\frac{1 }{\epsilon_{UV}} - 1 + \frac{\pi^2}{12} \epsilon\biggr).
\end{align}
Combing the virtual and real correction contributions shown in Eqs.~\eqref{eq:SGD-vir}
and~\eqref{eq:SGD-real}, we derive the $F_{[SS] \to Q\bar Q[SS]}$ before renormalization
\begin{align}
F^{NLO}_{[SS] \to Q\bar Q[SS]}(x,M_H,m_Q,\mu_f)\biggr\vert_{bare} =& \frac{\alpha_s }{4\pi} \biggr
(\frac{4\pi x^2 \mu^2 e^{-\gamma_E}}{M_H^2} \biggr)^{\epsilon}\biggr\{ N_c
\biggr[-\frac{2}{\epsilon_{UV}^2}\delta(1-x)
\nonumber\\
&\hspace{-4.5cm}
+ \frac{2}{\epsilon_{UV}}\biggr( \delta(1-x) + \frac{2x}{(1-x)_+}
\biggr)
- \biggr( \frac{4}{\epsilon_{UV}} - \frac{4}{\epsilon_{IR}} \biggr) \biggr( \frac{1}{2\Delta}\ln
\frac{1+\Delta}{1-\Delta} - 1 \biggr)\delta(1-x)
\nonumber\\
&\hspace{-4.5cm}
- \frac{4x}{(1-x)_+}- 8x  \biggr( \frac{\ln(1-x)}{1-x} \biggr)_+
\biggr]
+ \frac{1}{N_c} \biggr( \frac{2}{\epsilon_{UV}} - \frac{2}{\epsilon_{IR}}  \biggr) \biggr( \frac{1+
	\Delta^2}{2\Delta} \ln \frac{1+\Delta}{1-\Delta} -1 \biggr)
\nonumber\\
&\hspace{-4.5cm} \times
\delta(1-x)
- \biggr( \frac{1}{N_c} (1+\Delta^2)\frac{\pi^2}{\Delta} + N_c \frac{\pi^2}{6} \biggr) \delta(1-x) \biggr\}.
\end{align}
The ultraviolet divergences in the above result can be removed by
using the $\overline{\textrm{MS}}$ renormalization procedure, and we get the renormalized SGD
\begin{align}\label{eq:SGD-R}
F^{NLO}_{[SS] \to Q\bar Q[SS]}(x,M_H,m_Q,\mu_f) =& \frac{\alpha_s }{4\pi} \biggr
(\frac{4\pi x^2 \mu_c^2 e^{-\gamma_E}}{M_H^2} \biggr)^{\epsilon}
\biggr\{ N_c
\biggr[
\frac{4}{\epsilon_{IR}}  \biggr( \frac{1}{2\Delta}\ln
\frac{1+\Delta}{1-\Delta} - 1 \biggr)
\nonumber\\
&\hspace{-4.5cm}\times\delta(1-x)
- \frac{4x}{(1-x)_+}- 8x  \biggr( \frac{\ln(1-x)}{1-x} \biggr)_+ + 2\biggr( 3- \frac{1}{\Delta}\ln
\frac{1+\Delta}{1-\Delta}
\biggr)\delta(1-x)\ln\biggr(\frac{\mu_f^2}{M_H^2}\biggr)
\nonumber\\
&\hspace{-4.5cm}
-\delta(1-x)\ln^2\biggr(\frac{\mu_f^2}{M_H^2}\biggr) +\frac{4x}{(1-x)_+} \ln\biggr(\frac{x^2 \mu_f^2}{M_H^2}\biggr)
\biggr]
- \frac{2}{\epsilon_{IR}} \frac{1}{N_c}   \biggr( \frac{1+
	\Delta^2}{2\Delta} \ln \frac{1+\Delta}{1-\Delta} -1 \biggr)
\nonumber\\
&\hspace{-4.5cm} \times \delta(1-x) + \frac{1}{N_c}   \biggr( \frac{1+
	\Delta^2}{\Delta} \ln \frac{1+\Delta}{1-\Delta} -2 \biggr)\Delta(1-x)\ln\biggr(\frac{\mu_f^2}{M_H^2}\biggr)
-  \delta(1-x)
\nonumber\\
&\hspace{-4.5cm} \times
\biggr( \frac{1}{N_c} (1+\Delta^2)\frac{\pi^2}{\Delta} + N_c \frac{\pi^2}{6} \biggr) \biggr\}.
\end{align}
Here we distinguish the dimensional regularization scale $\mu_c$ from  the factorization scale $\mu_f$.

Similarly we can calculate the other SGDs by projecting the initial and finial $Q \bar Q$ pair into corresponding
states, and we have
\begin{subequations}\label{eq:SGDs}
	\begin{align}
	F_{[SD,T] \to Q\bar Q[SS]}^{NLO}(x,M_H,m_Q,\mu_f) \biggr \vert_{vir.}
	=& \frac{\alpha_s N_c}{4\pi} \sqrt{\frac{2(d-1)}{(d-2)(d+1)}} \frac{M_H^2}{36}
	\biggr( - \frac{1}{\epsilon_{IR}} +  \ln \frac{\mu_f^2}{4\pi
		\mu_c^2 e^{-\gamma_E}}\biggr)
	\nonumber\\
	&\hspace{-4.5cm}  \times \biggr( -3 + 2\Delta^2 + \frac{3(1-\Delta ^2 )}{2\Delta}
	\ln\frac{1+\Delta}{1-\Delta}\biggr)   \delta(1-x),
	\\
	F_{[SD,T] \to Q\bar Q[SS]}^{NLO}(x,M_H,m_Q,\mu_f) \biggr\vert_{real} =&
	\frac{\alpha_s N_c}{4\pi} \sqrt{\frac{2(d-1)}{(d-2)(d+1)}} \frac{4m_Q^2 (2- 3\mathcal {T} )+ M_H^2}{18}
	\nonumber\\
	&\hspace{-4.5cm}
	\times
	\biggr (\frac{4\pi x^2 \mu_c^2 e^{-\gamma_E}}{M_H^2} \biggr)^{\epsilon} \biggr[ -\frac{1}{2\epsilon_{IR}}\delta(1-x) - \frac{4m_Q^2 (9 \mathcal {T} - 8) - M_H^2}{ 6 (4m_Q^2 (2 - 3 \mathcal {T}) + M_H^2 )}\delta(1-x)
	\nonumber\\
	&\hspace{-4.5cm}
	+ \frac{x}{(1-x)_+}  \biggr] ,  \\
	F_{[DD,T] \to Q\bar Q[SS]}^{NLO}(x,M_H,m_Q,\mu_f) \biggr \vert_{vir.} =& 0,
	\\
	F_{[DD,T] \to Q\bar Q[SS]}^{NLO}(x,M_H,m_Q,\mu_f)  \biggr\vert_{real} =&
	\frac{\alpha_s N_c}{4\pi} \frac{2(d-1)}{(d-2)(d+1)}  \frac{(4m_Q^2 (2- 3\mathcal {T})+ M_H^2)^2}{108}
	\nonumber\\
	&\hspace{-4.5cm}
	\times
	\biggr (\frac{4\pi x^2 \mu_c^2 e^{-\gamma_E}}{ M_H^2} \biggr)^{\epsilon} \biggr[ -\frac{1}{2\epsilon_{IR}}\delta(1-x) + \frac{x}{(1-x)_+}
	\nonumber\\
	&\hspace{-4.5cm}
	- \frac{16 m_Q^4 (189 \mathcal {T}^2 - 204 \mathcal {T} + 52 ) - 8 m_Q^2 (87 \mathcal {T} - 50) M_H^2 +
		37 M_H^4 )}{ 24 (4m_Q^2 (2 - 3 \mathcal {T}) + M_H^2 )^2} \delta(1-x)   \biggr]
	.
	\end{align}
\end{subequations}
Here $\mathcal {T}$ is a hypergeometric function
\begin{align}
\mathcal {T} =\,_2F_1\left( \frac{1}{2} , 1 , \frac{3}{2} - \epsilon , \Delta^2  \right),
\end{align}
which can be expended as
\begin{align}
\mathcal {T} =& \frac{1}{2 \Delta}\biggr [  \ln \frac{1+\Delta}{1-\Delta} + \biggr( - 2\ln \frac{1+\Delta}{1-\Delta} -\textrm{Li}_2 \biggr( \frac{2\Delta}{ \Delta -1 } \biggr) + \textrm{Li}_2 \biggr( \frac{2\Delta}{ \Delta +1 } \biggr) \biggr) \epsilon \biggr]+O(\epsilon^2).
\end{align}

\section{Renormalization group equation and models for SGDs}\label{sec:RGE}

\subsection{Renormalization group equation}

Based the factorization formula in Eq.~\eqref{eq:fac1d-1}, RGEs for SGDs have the following general form
\begin{align}\label{eq:general-RGE-SGDs}
\frac{d}{d \ln \mu_f}  F_{[L^\prime \tilde{L}^\prime, \lambda^\prime] \to H}(z,M_H,m_Q, \mu_f)
=& \sum_{L, \tilde{L}, \lambda} \int_z^1 \frac{dx}{x} \boldsymbol{K}_{[L^\prime \tilde{L}^\prime, \lambda^\prime]}^{[L\tilde{L}, \lambda]}(\hat z, M_H/x, m_Q,\mu_f)
\nonumber\\
&\hspace{-4.5cm} \times F_{[L \tilde{L}, \lambda] \to H}(x,M_H,m_Q, \mu_f),
\end{align}
with $\hat{z}\equiv z/x$. The evolution kernel $\boldsymbol{K}$ can be perturbatively calculated by using the matching procedure. At LO in $\alpha_s$, we have
\begin{align}
\boldsymbol{K}_{[L^\prime \tilde{L}^\prime, \lambda^\prime]}^{[L\tilde{L}, \lambda],LO}(\hat z, M_H/x, m_Q,\mu_f)= &  \frac{d}{d \ln \mu_f}  F_{[L^\prime \tilde{L}^\prime, \lambda^\prime] \to Q\bar Q[L\tilde{L}, \lambda]}^{NLO}(\hat z,M_H/x,m_Q, \mu_f).
\end{align}
For simplicity, in the following we only consider the evolution equation for $F_{[SS] \to H}$ and ignore contributions from $F_{[SD,T] \to H}$ and $F_{[DD,T] \to H}$ because they are $v^4$ suppressed.

From Eq.~\eqref{eq:SGD-R}, we have the evolution kernel
\begin{align}
\boldsymbol{K}_{[SS]}^{[SS],LO}(z, M_H, m_Q,\mu_f) =& \frac{\alpha_s }{\pi} \biggr \{ N_c \biggr [\frac{2z}{(1-z)_+} -\ln \frac{\mu^2 e^{-1}}{M_H^2}
\delta(1-z)
\nonumber\\
&\hspace{-3cm}
- 2\delta(1-z)\biggr( \frac{1}{2\Delta}\ln \frac{1+\Delta}{1-\Delta} - 1 \biggr)
\biggr ]  + \frac{1}{N_c} \biggr( \frac{1+ \Delta^2}{2\Delta} \ln \frac{1+\Delta}{1-\Delta} -1 \biggr)\delta(1-z) \biggr \}.
\end{align}
The evolution equation with this kernel is difficult to be solved analytically. However, if we only want to resum large logarithms at leading order in $v^{2}$ expansion, we only need to keep
the leading power term in the evolution kernel, which reads
\begin{align}\label{eq:SGD-kernel}
\boldsymbol{K}_{[SS]}^{[SS],LO(0)}( z, M_H, \mu_f)= &\frac{\alpha_s }{4\pi} \biggr[ \Gamma_0^F \biggr ( \frac{ z}{(1- z)_+}  -\ln \frac{ \mu_f }{M_H}
\delta(1- z)\biggr ) -  \gamma^F_0 \delta(1- z) \biggr ],
\end{align}
where
\begin{align}\label{eq:cusp}
\Gamma^F_0 = 8 N_c,  \quad \gamma^F_0 =& -4 N_c .
\end{align}
Then the evolution equation for $F_{[SS] \to H}$ becomes
\begin{align}\label{eq:RGE-SGD}
\frac{d}{d \ln \mu_f}  F_{[SS] \to H}(z,M_H,m_Q, \mu_f)
=& \int_z^1 \frac{dx}{x} \boldsymbol{K}_{[SS]}^{[SS],LO(0)}(\hat z, M_H/x,\mu_f)  \nonumber\\
& \times F_{[SS] \to H}(x,P_H,m_Q, \mu_f).
\end{align}

To solve the above evolution equation, we rewrite it as a function of the variable $\omega=M_H(1/z-1)$, which results in
\begin{align}\label{eq:RGE-SGD1}
& \frac{d}{d \ln\mu_f}
F_{[SS] \to H}(\omega,M_H,m_Q, \mu_f)
\nonumber\\
&= \int_0^\omega d\omega'  \frac{\alpha_s}{4\pi} \biggr [\Gamma^F_0 \biggr ( \biggr[ \frac{1}{\omega-\omega'} \biggr]_+ - \ln \mu_f
\delta(\omega-\omega')\biggr ) -\gamma^F_0 \delta(\omega - \omega^\prime) \biggr]
\nonumber\\
& \quad \times
F_{[SS] \to H}(\omega^\prime,M_H,m_Q, \mu_f),
\end{align}
with $\omega^\prime = M_H(1/x-1)$.
To derive the above equation, we have used the rescaling identity for plus functions~\cite{Fleming:2007xt}
\begin{align}
\kappa \biggr[ \frac{\ln^n(\kappa\omega)}{\kappa \omega}\biggr]_+
=&  \sum_{i=0}^n \frac{n!}{(n-i)! i!} \ln^{n-i}(\kappa) \biggr[ \frac{\ln^i(\omega)}{ \omega}\biggr]_+ + \frac{\ln^{n+1}(\kappa)}{n+1}\delta(\omega) .
\end{align}
The plus functions of the dimensionful variable $\omega$ are defined as
\begin{align}
\int_0^\xi d\omega \biggr[ \frac{\ln^n(\omega)}{\omega}\biggr]_+ f(\omega)
=&  \int_0^\xi d\omega  \frac{\ln^n(\omega)}{\omega} \biggr( f(\omega)-f(0)\biggr)+ \frac{1}{n+1} \ln^{n+1}(\xi)f(0),
\end{align}
where $f(\omega)$ is a smooth test function.
To solve the RGE Eq.~\eqref{eq:RGE-SGD1}, it is convenient to perform a Laplace transformation~\cite{Becher:2006nr,Becher:2006mr,Becher:2007ty}, which, together with its inverse, is given by
\begin{equation}
\tilde{f}(s)= \int_0^\infty d\omega e^{- \omega s}f(\omega) , \quad \textrm{and} \quad f(\omega)= \frac{1}{2\pi i}\int_{c-i \infty}^{c + i\infty} ds e^{ \omega s}\tilde{f}(s),
\end{equation}
where the constant $c$ is chosen to be larger than the real part of the rightmost singularity of $\tilde{f}(s)$. Taking a Laplace transform in Eq.~\eqref{eq:RGE-SGD1} we have a simple multiplicative RGE
\begin{align}\label{eq:LaplaceRGE}
\frac{d}{d \ln\mu_f}
\tilde F_{[SS] \to H}(s , M_H,m_Q, \mu_f)
=&  -\frac{\alpha_s}{4\pi} \biggr(  \Gamma^F_{0}   \ln (\bar s \mu_f )
+ \gamma^F_0 \biggr) \nonumber\\
&  \times \tilde F_{[SS] \to H}(s , M_H,m_Q, \mu_f),
\end{align}
where $\bar s=s e^{\gamma_E}$.

Solving the RGE in Laplace space is straight forward, details of which are given in App.~\ref{sec:app-rge}. We eventually obtain
\begin{align}\label{eq:Resummed-SGD}
\tilde {F}_{[SS] \to H}(s,M_H,m_Q,  M_H) =&    \mathrm{exp}\biggr[ h_0(\chi^\ast) \ln(M_H \bar s)  + h_1(\chi^\ast) \biggr] \tilde F_{\textrm{fix}}(M_H) \tilde F^{\mathrm{mod}}(s),
\end{align}
where $\tilde F_{\textrm{fix}}(M_H)$ is given in Eq.~\eqref{eq:fixorder} which recovers fixed-order perturbative results at small $s$ region, functions $h_{0,1}$ are given by Eq.~\eqref{eq:resummed-F},
and $\chi^\ast$ is defined as
\begin{align}
\chi^\ast = &\frac{\beta_0}{2\pi}  \alpha_s(M_H) \ln (M_H \bar s) + \frac{\beta_0}{2\pi}  \alpha_s(M_H)\ln \biggr( 1+ \frac{1}{M_H \bar s^*} \biggr)
\nonumber\\
& - \frac{\beta_0}{2\pi}  \alpha_s(M_H)\ln \biggr( 1+ \frac{\bar s}{\bar s^*} \biggr).
\end{align}
with
\begin{align}
	\beta_0=& \frac{11N_c-2 n_f}{6},
\end{align}
where $n_f$ represents the number of light flavors. We have introduced a ``frozen'' scale $\bar s^*$ with
\begin{align}
\bar s^*  =& \frac{a}{M_H} \mathrm{exp}\biggr[ \frac{\pi}{\beta_0 \alpha_s(M_H)}\biggr]
\end{align}
to prevent Landau singularity,
where $a$ (with $a \gtrsim 1$) is a parameter and its dependence should be compensated with a change of nonperturbative input. Therefore, we fix it as $a=1.3$ in the rest of this paper. Finally all nonperturbative information is included in the model $\tilde F^{\textrm{mod}}(s)$. Then SGD can be obtained by transforming Eq.~\eqref{eq:Resummed-SGD} back to momentum space using
\begin{align}\label{eq:MResummed-SGD}
F_{[SS] \to H} (x,M_H,m_Q,  M_H) =&    \frac{1}{2\pi i}\int_{c-i \infty}^{c + i\infty} ds e^{ M_H(1/x-1) s} \tilde {F}_{[SS] \to H}(s,M_H,m_Q,  M_H),
\end{align}
where $c$ is a constant that must be greater than the real part of all singularities of $\tilde {F}_{[SS] \to H}$.

\subsection{Sensitivity of nonperturbative model}

We will demonstrate that SGD obtained above by solving RGE is only sensitive to two parameters of the nonperturbative model. To this end, we will study various models, with significant different shapes. We choose the model function used in \cite{Fleming:2003gt,Fleming:2006cd,Bauer:2001rh,Sun:2018yam} as our first class of models,
\begin{align}\label{eq:model-SGD}
F^{\textrm{mod}}(\omega^\prime ) =& M_H N_H \frac{b^{b}}{\Gamma(b)} \frac{\omega^{\prime b-1}}{\bar{\Lambda}^{b}} e^{-b\omega^\prime/\bar{\Lambda}} ,  \quad \omega^\prime  = M_H(1/x-1),
\end{align}
whose zeroth, first and second moments are $M_HN_H$, $M_HN_H  \bar{\Lambda}$ and $M_HN_H  \bar{\Lambda}^2(\frac{1}{b}+1)$, respectively.
Here $N_H$ determines the normalization,  $\bar{\Lambda}$ characterizes the average radiated momentum, and $b$ is related to the width of the model function. We vary $\bar{\Lambda}$ from $0.5\mathrm{GeV} - 0.7\mathrm{GeV}$ \footnote{There is yet no first-principle way to determine  $\bar{\Lambda}$, but we can guess its order of magnitude. For the transition from $\COcSa$
to a $1^{--}$ quarkonium, where a double color E1 transition should dominate, a good guess is that  $\bar{\Lambda}$ is at the order of mass splitting between 2S state and 1S state. Due to $M_{\psi(2S)}-M_{J/\psi} \approx 0.59 \mathrm{GeV}$ and  $M_{\Upsilon(2S)}-M_{\Upsilon(1S)} \approx 0.56 \mathrm{GeV}$, we choose the central value of $\bar{\Lambda}$ at $0.6\mathrm{GeV}$, although it is adjustable.} and $b$ from $1 - 3$. Other models are chosen to cover various possible shapes, but with the same normalization ($M_H N_H$) and average radiated momentum  ($0.6\mathrm{GeV}$). All models that we will study are listed in the following, 
\begin{align}\label{eq:models}
&\textrm{Model-1:}~~F^{\textrm{mod}}(\omega^\prime )\vert_{\bar{\Lambda}=0.6\mathrm{GeV},b=2},
\quad\quad \textrm{Model-2:}~~F^{\textrm{mod}}(\omega^\prime )\vert_{\bar{\Lambda}=0.6\mathrm{GeV},b=1},
\nonumber\\
&\textrm{Model-3:}~~F^{\textrm{mod}}(\omega^\prime )\vert_{\bar{\Lambda}=0.6\mathrm{GeV},b=3},
\quad\quad
\textrm{Model-4:}~~F^{\textrm{mod}}(\omega^\prime )\vert_{\bar{\Lambda}=0.5\mathrm{GeV},b=2},
\nonumber\\
&\textrm{Model-5:}~~F^{\textrm{mod}}(\omega^\prime )\vert_{\bar{\Lambda}=0.7\mathrm{GeV},b=2},\nonumber\\
&\textrm{Model-6:}~~4M_HN_H[\theta(w^\prime\geq \frac{19}{40})-\theta(w^\prime > \frac{29}{40})],
\nonumber\\
& \textrm{Model-7:}~~\frac{5}{6}M_HN_H[\theta(w^\prime\geq 0)-\theta(w^\prime > \frac{6}{5})],
\nonumber\\
&\textrm{Model-8:}~~\begin{cases}
M_HN_H(-\frac{50}{81}w^\prime + \frac{10}{9}), \quad 0\leq w^\prime \leq \frac{9}{5},\\
0, \quad  w^\prime >\frac{9}{5},
\end{cases},
\nonumber\\
&\textrm{Model-9:}~~\begin{cases}
\frac{200}{81}M_HN_Hw^\prime , \quad 0\leq w^\prime \leq \frac{9}{10},\\
0, \quad  w^\prime >\frac{9}{10}.
\end{cases}
\end{align}

\begin{figure}[htb!]
	\begin{center}
		\vspace*{0.8cm}
		\hspace*{-5mm}
		\includegraphics[width=0.45\textwidth]{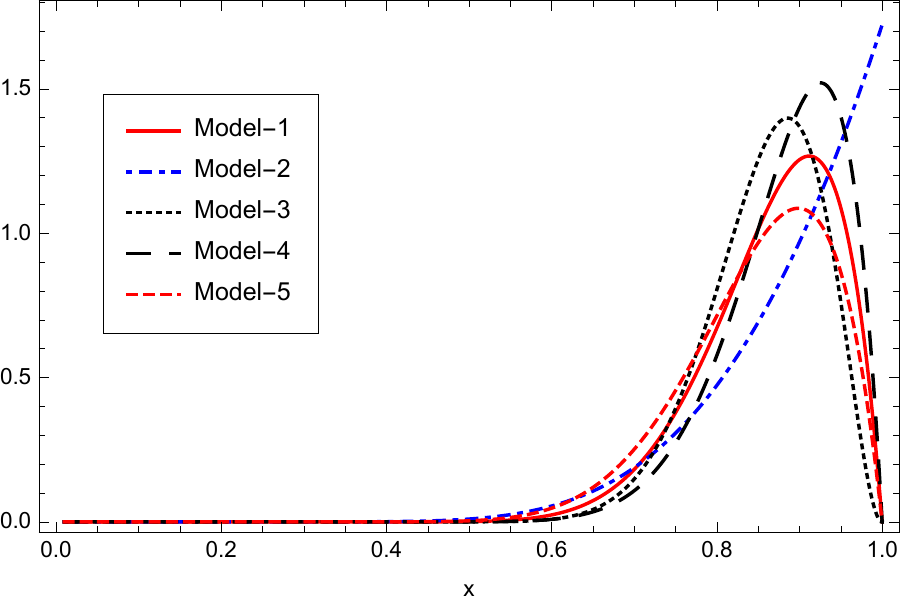}
		\hspace*{5mm}
		\includegraphics[width=0.45\textwidth]{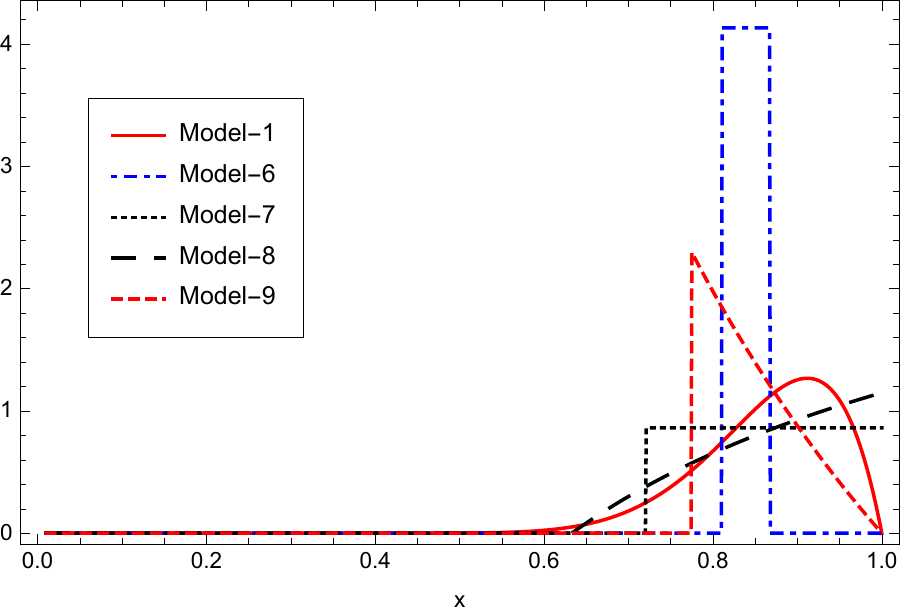}
	\end{center}
	\begin{center}
		\vspace*{0.8cm}
		\hspace*{-5mm}
		\includegraphics[width=0.45\textwidth]{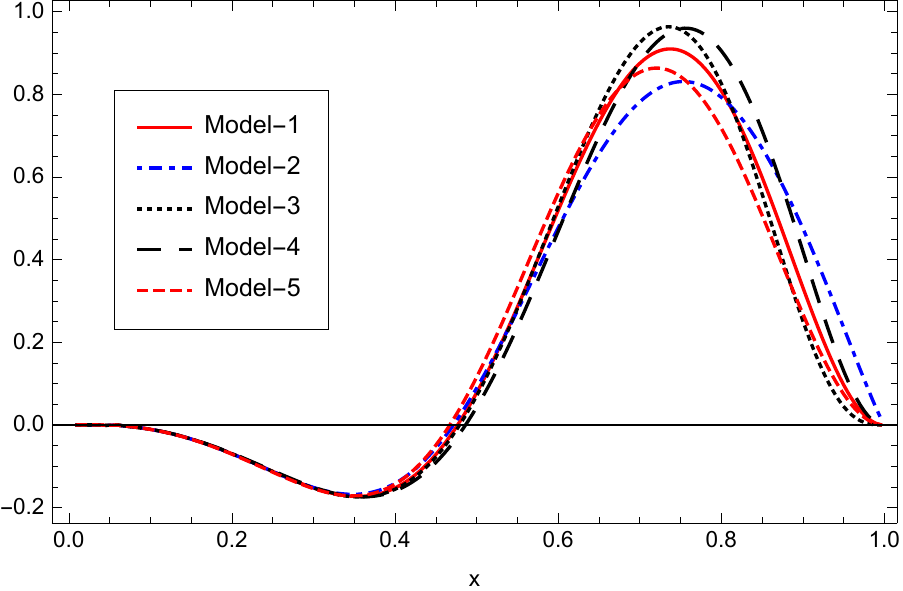}
		\hspace*{5mm}
		\includegraphics[width=0.45\textwidth]{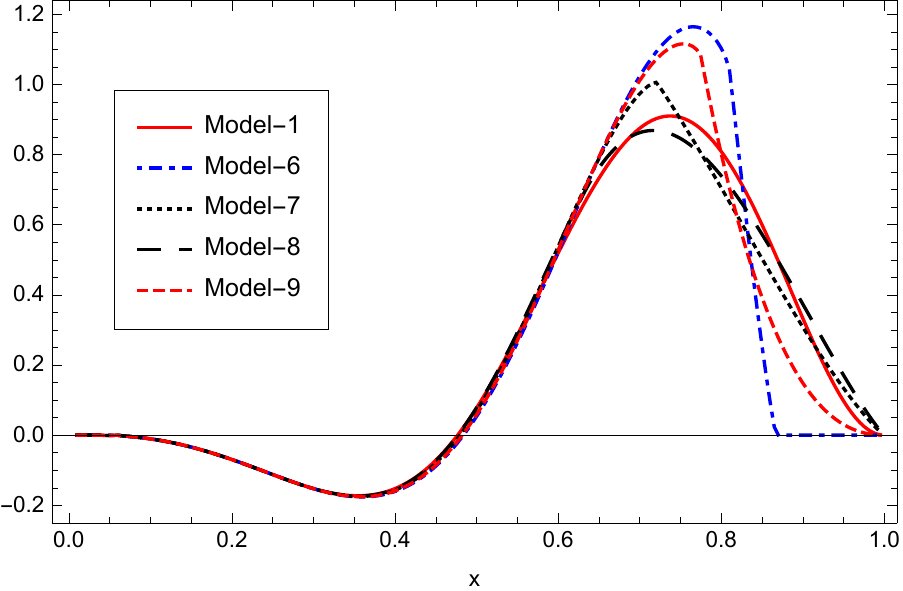}
	\end{center}
	\vspace*{-.5cm}
	\caption{The first row shows the different models, the corresponding SGDs at LO
             are given in the second row. \label{fig:model-SGD}}
	\vspace*{0.cm}
\end{figure}

\begin{table}
	\caption[]{The differences between the Models and the SGDs .}\label{table:difference}
	\begin{center}
		\begin{tabular}{|c|c|c|c|c|c|}
			\hline
			\multicolumn{6}{|c|}{Models}\\
			\hline
			\multirow{1}{*}{$\langle g_1, g_2 \rangle$} &\multirow{1}{*}{\textrm{Model-1}}&\multirow{1}{*}{\textrm{Model-2}}
			&\multirow{1}{*}{\textrm{Model-3}}&\multirow{1}{*}{\textrm{Model-4}}
			&\multirow{1}{*}{\textrm{Model-5}} \\
			\hline
			\multirow{1}{*}{Model-1}& 0 & 0.190  & 0.107 & 0.093 & 0.078  \\
			\hline
			\multirow{1}{*}{Model-2}&  & 0  & 0.291 & 0.191 & 0.210  \\
			\hline
			\multirow{1}{*}{Model-3}&  &   & 0 & 0.160 & 0.122  \\
			\hline
			\multirow{1}{*}{Model-4}&  &   &  & 0 & 0.170  \\
			\hline
			\multirow{1}{*}{Model-5}&  &   &  &  & 0  \\
			\hline
		\end{tabular}
	\end{center}
	\begin{center}
		\begin{tabular}{|c|c|c|c|c|c|}
			\hline
			\multicolumn{6}{|c|}{SGDs}\\
			\hline
			\multirow{1}{*}{$\langle g_1, g_2 \rangle$} &\multirow{1}{*}{\textrm{Model-1}}&\multirow{1}{*}{\textrm{Model-2}}
			&\multirow{1}{*}{\textrm{Model-3}}&\multirow{1}{*}{\textrm{Model-4}}
			&\multirow{1}{*}{\textrm{Model-5}} \\
			\hline
			\multirow{1}{*}{Model-1}& 0 & 0.060  & 0.029 & 0.048 & 0.045  \\
			\hline
			\multirow{1}{*}{Model-2}&  & 0  & 0.089 & 0.061 & 0.083  \\
			\hline
			\multirow{1}{*}{Model-3}&  &   & 0 & 0.062 & 0.049  \\
			\hline
			\multirow{1}{*}{Model-4}&  &   &  & 0 & 0.093  \\
			\hline
			\multirow{1}{*}{Model-4}&  &   &  &  & 0  \\
			\hline
		\end{tabular}
	\end{center}
	\begin{center}
		\begin{tabular}{|c|c|c|c|c|c|}
			\hline
			\multicolumn{6}{|c|}{Models}\\
			\hline
			\multirow{1}{*}{$\langle g_1, g_2 \rangle$} &\multirow{1}{*}{\textrm{Model-1}}&\multirow{1}{*}{\textrm{Model-6}}
			&\multirow{1}{*}{\textrm{Model-7}}&\multirow{1}{*}{\textrm{Model-8}}
			&\multirow{1}{*}{\textrm{Model-9}} \\
			\hline
			\multirow{1}{*}{Model-1}& 0 & 0.777  & 0.204 & 0.144 & 0.366  \\
			\hline
			\multirow{1}{*}{Model-6}&  & 0  & 0.794 & 0.818 & 0.633  \\
			\hline
			\multirow{1}{*}{Model-7}&  &   & 0 & 0.146 & 0.371  \\
			\hline
			\multirow{1}{*}{Model-8}&  &   &  & 0 & 0.427 \\
			\hline
			\multirow{1}{*}{Model-9}&  &   &  &  & 0  \\
			\hline
		\end{tabular}
	\end{center}
	\begin{center}
		\begin{tabular}{|c|c|c|c|c|c|}
			\hline
			\multicolumn{6}{|c|}{SGDs}\\
			\hline
			\multirow{1}{*}{$\langle g_1, g_2 \rangle$} &\multirow{1}{*}{\textrm{Model-1}}&\multirow{1}{*}{\textrm{Model-6}}
			&\multirow{1}{*}{\textrm{Model-7}}&\multirow{1}{*}{\textrm{Model-8}}
			&\multirow{1}{*}{\textrm{Model-9}} \\
			\hline
			\multirow{1}{*}{Model-1}& 0 & 0.129  & 0.041 & 0.029 & 0.084  \\
			\hline
			\multirow{1}{*}{Model-6}&  & 0  & 0.119 & 0.148 & 0.056  \\
			\hline
			\multirow{1}{*}{Model-7}&  &   & 0 & 0.036 & 0.065  \\
			\hline
			\multirow{1}{*}{Model-8}&  &   &  & 0 & 0.098  \\
			\hline
			\multirow{1}{*}{Model-9}&  &   &  &  & 0  \\
			\hline
		\end{tabular}
	\end{center}
\end{table}

These input models are shown in the first row in Fig.~\ref{fig:model-SGD}, and the corresponding SGDs obtained by solving RGE at LO are given in the second row, where we have taken $M_H=3.1 \mathrm{GeV}$, $n_f = 3$, $\Lambda_{QCD}^{(3)}= 0.25 \mathrm{GeV}$ and $N_H=1/3$. From the plots we find that, comparing with input models, SGDs are much broader and their peaks are shifted to smaller $x$. More importantly, SGDs at moderate and small $x$, i.e., $x<0.65$, are almost independent of input models, as far as the models have the same normalization and average radiated momentum. This is reasonable because small-$x$ region is dominated by perturbative effects. At larger $x$ region, values of SGDs depend on input models, but the sensitivity after RGE running is much weaker than the original models. To see this more clearly, we introduce a function to describe the difference between two functions $g_1(x)$ and $g_2(x)$
\begin{align}
\langle g_1, g_2 \rangle=\frac{\int_0^1|g_1(x)-g_2(x)|dx}{\int_0^1(|g_1(x)|+|g_2(x)|)dx}.
\end{align}
Differences between input models and that between corresponding SGDs are shown in Table.~\ref{table:difference}.
It is clear that differences are significantly decreased after performing resummation using RGE. Especially for models with the same average radiated momentum, differences have been typically reduced by a factor of $1/5$. 

The above phenomenon can be interpreted as following. The overall normalization of a SGD is fully determined by the normalization of the input model. The shape of a SGD is almost fully determined by perturbative calculation with all-order resummation, although its peak can be affected by the average radiated momentum of the input model. In this consideration, in practical use it is sufficient to choose a suitable parameter $b$ in Eq.~\eqref{eq:model-SGD}, and fit the parameters $N_H$ and $\bar{\Lambda}$ by comparing with experiment data.

\section{Gluon fragmentation function of quarkonium in SGF}\label{sec:FF-SGF}

\subsection{SGF formula for gluon fragmentation function }
The Collins-Soper definition for fragmentation function of a gluon fragmenting into a hadron (quarkonium) $H$ is given by
\cite{Collins:1981uw}
\begin{align}\label{eq:defFF}
    D_{g \rightarrow H}(z,\mu)=
    & \frac{-g_{\mu\nu}z^{d-3}}{ P_c^{+}(N_{c}^{2}-1)(d-2)} \int \frac{d\xi^{-}}{2 \pi} e^{-i
    P_c^{+} \xi^{-}}  \langle 0 | G_{c}^{+\mu}(0) \Phi_l^{\dag}(0)_{cb}
    \mathcal{P}_{H(P_H)} \nonumber
    \\
    & \times \Phi_l(\xi^{-})_{ba}  G_{a}^{+\nu}(0,\xi^{-},0_{\perp}) | 0
    \rangle ,
\end{align}
where $G^{\mu\nu}$ is the gluon field-strength operator, $P_H$ and $P_c$ are the momenta of the hadron and
initial virtual gluon, respectively, and $z$ is the ``$+$'' momentum fraction defined as $z=P_H^+/P_c^+$.
The projection operator $\mathcal{P}_{H(P_H)}$ is given by
\begin{equation}\label{eq:projectH}
\mathcal{P}_{H(P_H)} = \sum_X |H(P_H)+X \rangle \langle H(P_H)+X|\,,
\end{equation}
where $X$ sums over all unobserved particles.
The gauge link $\Phi_l(\xi^{-})$ is the same as the one given in Eq.~\eqref{eq:gaugelink}. The definition of Eq.~\eqref{eq:defFF} is gauge invariant and we use the Feynman gauge in our calculation.

In Eq.~\eqref{eq:defFF}, $\mu$ is the collinear factorization scale. The dependence of
FFs on $\mu$ is controlled by the DGLAP evolution equations~\cite{Gribov:1972ri,Altarelli:1977zs,Dokshitzer:1977sg},
 \begin{equation}\label{eq:DGLAP}
 \frac{d }{d\ln\mu} D_{g \rightarrow H}(z,\mu) = \int_z^1 \frac{dx}{x} P_{gg}(z/x,\mu)  D_{g \rightarrow H}(x,\mu),
\end{equation}
where  we have ignored contributions of quark fragmentation functions because they are usually less important. Evolving FFs from a low scale $\mu $ up to a high scale $\mu_h $ by solving the DGLAP evolution, one can resum large logarithms of $\mu^2/\mu_h^2$ to all orders in perturbation theory.
The gluon splitting function $P_{gg}$ has the perturbative expansion
\begin{align}\label{eq:splitting-function}
P_{gg}(x,\mu)=& \frac{\alpha_s(\mu)}{2\pi}P^{LO}_{gg}(x) + \biggr( \frac{\alpha_s(\mu)}{2\pi} \biggr)^2 P^{NLO}_{gg}(x) +  \cdots,
\end{align}
where, e.g., $P^{LO}_{gg}(x)$ is given by
\begin{align}
P^{LO}_{gg}(x)=& 2N_c\biggr[ \frac{x}{(1-x)_+} + \frac{1-x}{x} +x(1-x) + \frac{\beta_0}{2N_c} \delta(1-x)\biggr].
\end{align}

In the parton (initial virtual gluon) frame, the gluon FF to quarkonium has the form
\begin{align}
D_{g \to H}(z,M_H,m_Q,\mu) = \int \frac{d P_H^+ d^{2} P_{H\perp}}{(2\pi)^{3}2P_H^+} \delta(z-\frac{P_H^+}{P_c^+}) (2\pi)^{3} 2 P_H^0 \frac{d\sigma_{H}(P_c,P_H,m_Q^2)}{d^{3}P_{H}}.
\end{align}
Using the SGF-1d formula Eq.~\eqref{eq:fac1d-2} and changing the variables from the parton to the hadron frame, we have
\begin{align}
D_{g \to H}(z,M_H,m_Q,\mu) =& \int \frac{d P_H^+ d^{2} P_{H\perp}}{(2\pi)^{3}2P_H^+} \delta(z-\frac{P_H^+}{P_c^+})
\sum_{n,n^\prime} \int \frac{dx}{x^2}   d\hat{\sigma}_{[nn^\prime]}(P_H/x ,m_Q, \mu_f)
 \nonumber\\
&\times F_{[nn^\prime] \to H }(x, M_H, m_Q, \mu_f) \nonumber\\ &\hspace{-2cm}
= \sum_{n,n^\prime} \int \frac{dx}{x}  \frac{z}{x}P_c^+\int \frac{d P_c^+ d^{2} P_{c\perp}}{(2\pi)^{3}2P_c^+} \delta(P_c^+-\frac{P_H^+/x}{z/x})  d\hat{\sigma}_{[nn^\prime]}(P_H/x, m_Q,\mu_f)
\nonumber\\
& \hspace{-2cm} \quad   \times F_{[nn^\prime] \to H }(x, M_H, m_Q, \mu_f).
\end{align}
By
defining the short
distance hard parts $\hat{D}_{[nn^\prime]}(\hat{z},M_H/x, m_Q,\mu, \mu_f)$ as following
\begin{align}
\hat{D}_{[nn^\prime]}(\hat{z},M_H/x, m_Q,\mu, \mu_f) =& \frac{z}{x}P_c^+\int \frac{d P_c^+ d^{2} P_{c\perp}}{(2\pi)^{3}2P_c^+} \delta(P_c^+-\frac{P_H^+/x}{z/x})  d\hat{\sigma}_{[nn^\prime]}(P_H/x,m_Q,\mu_f),
\end{align}
we obtain
\begin{align}\label{eq:general-FFSGF-cut}
D_{g \to H}(z,M_H,m_Q,\mu) =&  \sum_{n,n^\prime} \int \frac{dx}{x}  \hat{D}_{[nn^\prime]}(\hat{z},M_H/x,m_Q, \mu, \mu_f)
 F_{[nn^\prime] \to H}(x,M_H,m_Q, \mu_f),
\end{align}
which is the general SGF formula for heavy quarkonium FF. Such a factorization formula holds for any quarkonium state $H$, and the $H$ dependence only appears in SGDs.

Then for the purpose of this paper, where we are only interested in $\state{{3}}{S}{1,\lambda}{8}$ and $\state{{3}}{D}{1,\lambda}{8}$ states, we have
\begin{align}\label{eq:FFSGF-1}
D_{g \to H}&(z,M_H,m_Q,\mu) =  \sum_{\lambda=T,L} \int_{z}^1 \frac{dx}{x}  \nonumber\\
&\biggr[\hat{D}_{[SS]}(\hat{z},M_H/x, m_Q, \mu, \mu_f)\times  F_{[SS] \to H}(x,M_H,m_Q, \mu_f)  \nonumber\\
& + 2\hat{D}_{[SD,\lambda]}(\hat{z},M_H/x, m_Q, \mu, \mu_f)\times  F_{[SD,\lambda] \to H}(x,M_H,m_Q, \mu_f)  \nonumber\\
& + \hat{D}_{[DD,\lambda]}(\hat{z},M_H/x, m_Q,\mu, \mu_f)\times F_{[DD,\lambda] \to H}(x,M_H,m_Q, \mu_f) \biggr],
\end{align}
where we have use the fact that polarized SGDs for $S$-wave states can be related polarization-summed SGDs, as noted in Eq.~\eqref{eq:reduce}.

Similar to Eq.~\eqref{eq:matchingEq1}, we replace the quarkonium $H$ in Eq.~\eqref{eq:FFSGF-1} by an on-shell $Q \bar Q$ pair with specific quantum number and momenta $P_H$.
Then we have following matching relations
\begin{align}\label{eq:matchingEq3}
\hat{D}_{[SS]}^{LO}(\hat z, M_H/x, m_Q, \mu, \mu_f) =&  D^{LO}_{g \to Q\bar Q[SS] }(\hat z, M_H/x, m_Q,\mu) ,
\nonumber\\
\hat{D}_{[SD,\lambda]}^{LO}(\hat z, M_H/x, m_Q, \mu, \mu_f) =&  D^{LO}_{g \to Q\bar Q[SD,\lambda] }(\hat z, M_H/x, m_Q,\mu) ,
\nonumber\\
\hat{D}_{[DD,\lambda]}^{LO}(\hat z, M_H/x, m_Q, \mu, \mu_f) =&  D^{LO}_{g \to Q\bar Q[DD,\lambda] }(\hat z, M_H/x, m_Q,\mu),
\nonumber\\
\hat{D}_{[SS]}^{NLO}(\hat z, M_H/x ,m_Q, \mu, \mu_f) =&  D^{NLO}_{g \to Q\bar Q[SS] }(\hat z, M_H/x, m_Q,\mu) - \sum_{\lambda=T,L}\int_{0}^1 \frac{dy}{y}
\nonumber\\
&\hspace{-4.2cm} \times \biggr[ \hat{D}_{[SS]}^{LO}(\hat z/y,M_H/(xy), m_Q, \mu, \mu_f)\times  F_{[SS] \to Q\bar Q[SS]}^{NLO}(y,M_H/x, m_Q, \mu_f)\nonumber\\
&\hspace{-4cm} + 2\hat{D}_{[SD,\lambda]}^{LO}(\hat z/y,M_H/(xy), m_Q, \mu, \mu_f) \times F_{[SD,\lambda] \to Q\bar Q[SS]}^{NLO}(y,M_H/x, m_Q, \mu_f) \nonumber\\
&\hspace{-4cm}  + \hat{D}_{[DD,\lambda]}^{LO}(\hat z/y,M_H/(xy), m_Q, \mu, \mu_f)\times F_{[DD,\lambda] \to Q\bar Q[SS]}^{NLO}(y,M_H/x, m_Q, \mu_f) \biggr] .
\end{align}
These relations enable us to calculate $\hat{D}_{[SS]}$ perturbatively to NLO.

\subsection{Perturbative calculation of gluon fragmentation functions}
\label{sec:perturbative-FF}

For the fragmentation functions we have
\begin{align}\label{eq:FF0}
\begin{split}
	D_{g\to Q\bar Q[L\tilde{L},\lambda]}(z,M_H, m_Q, \mu) =&
\frac{-z^{d-3}P_c^+}{(N_c^2-1)(d-2)} \int d \Phi
\mathcal{M}_{L\rho}(P_H,P_c,m_Q,k_i,\lambda) \\
\times &
\mathcal{M}_{\tilde{L}\sigma}^{\ast}(P_H,P_c,m_Q,k_i,\lambda)
( g^{\mu \rho}- \frac{P_H^\mu l^\rho}{P_c
\cdot l})
( g^{\sigma}_\mu - \frac{P_{H\mu} l^\sigma}{P_c \cdot l}) ,
\end{split}
\end{align}
where $d \Phi$ is the final state phase space, $k_i$ denotes momenta of the
final-state gluons, and $\mathcal{M}$ is projected
amplitude using projector
introduced in Eq.~\eqref{eq:spinor-operator}, i.e.
\begin{align}\label{eq:AMP0}
 \mathcal{M}_{L}^{\rho}(P_H,P_c,m_Q,k_i,\lambda)  =&    \int \frac{d^{d-2}\Omega}
 {N_\Omega} \textrm{Tr}\biggr[ \mathcal{A}^{\rho}_{g\to Q\bar Q  }
 \Pi[\state{{3}}{L}{1,\lambda}{8}] \biggr] ,
\end{align}
where $\mathcal{A}^{\rho}_{g\to Q\bar Q}$ is the amplitude from a virtual gluon to a $Q\bar Q$ pair plus light partons with spinors of $Q\bar Q$ removed, $\Omega$ is the solid angle of relative momentum $\boldsymbol{q}$ in the $Q\bar Q$ rest frame, and $N_\Omega$ is given by
\begin{align}
N_\Omega = \int d^{d-2}\Omega.
\end{align}

At LO in $\alpha_s$, the phase space $d \Phi^{LO}$ is given by
\begin{align}\label{eq:phase-0}
  d \Phi^{LO} &= 	
    \frac{d^{d} P_{c}}{(2\pi)^{d}}
    (2\pi)^{d} \delta^{d} \left(
    P_{c} - P_H \right) \delta \left( P_c^+ - \frac{P_H^{+}}{z} \right) \nonumber\\
    &= \frac{z}{P_H^+}\delta(1-z).
\end{align}
 Then the calculation of perturbative FFs is straightforward, which gives
\begin{subequations}\label{eq:FFLO}
\begin{align}
D^{LO}_{g \to Q\bar Q[SS]}(z, M_H,m_Q, \mu) =&  \bar D^{LO}(M_H,m_Q, \mu)\delta(1-z), \\
D^{LO}_{g \to Q\bar Q [SD,T]}(z,M_H,m_Q,\mu) =& D^{LO}_{g \to Q\bar Q [DS,T]}(z,M_H,m_Q,\mu) \nonumber\\
&\hspace{-4cm}= \frac{-2\sqrt{2(d-1)(d+1)(d-2)}}
{(M_H+2m_Q)(M_H(d-2)+2m_Q)} \bar D^{LO}(M_H,m_Q, \mu)\delta(1-z), \\
D^{LO}_{g \to Q\bar Q[DD,T]}(z,M_H,m_Q,\mu)& \nonumber\\
&\hspace{-4cm}= \frac{8(d-2)(d+1)(d-1)}
{(M_H+2m_Q)^2(M_H(d-2)+2m_Q)^2} \bar D^{LO}(M_H,m_Q, \mu)\delta(1-z) ,\\
D^{LO}_{g \to Q\bar Q [SD,L]}(z,M_H,m_Q,\mu)=&D^{LO}_{g \to Q\bar Q [DD,L]}(z,M_H,m_Q,\mu) \nonumber\\
&\hspace{-4cm} = 0,
\end{align}
\end{subequations}
where
\begin{align}
\bar D^{LO}(M_H,m_Q, \mu)\equiv \frac{8\pi \alpha_s
\mu^{2\epsilon}}{(N_c^2-1)M_H^3}\biggr (  \frac{M_H(d-2) + 2m_Q}{(d-1)M_H} \biggr)^2,
\end{align}
As we have mentioned before, basing on Eqs.~\eqref{eq:FFLO} and~\eqref{eq:matchingEq3}, we find the longitudinally polarized ($\lambda=L$) SGDs are irrelevant for calculating $\hat{D}_{[SS]}^{NLO}$.
\begin{figure}[htb!]
	\begin{center}
		\includegraphics[width=1\textwidth]{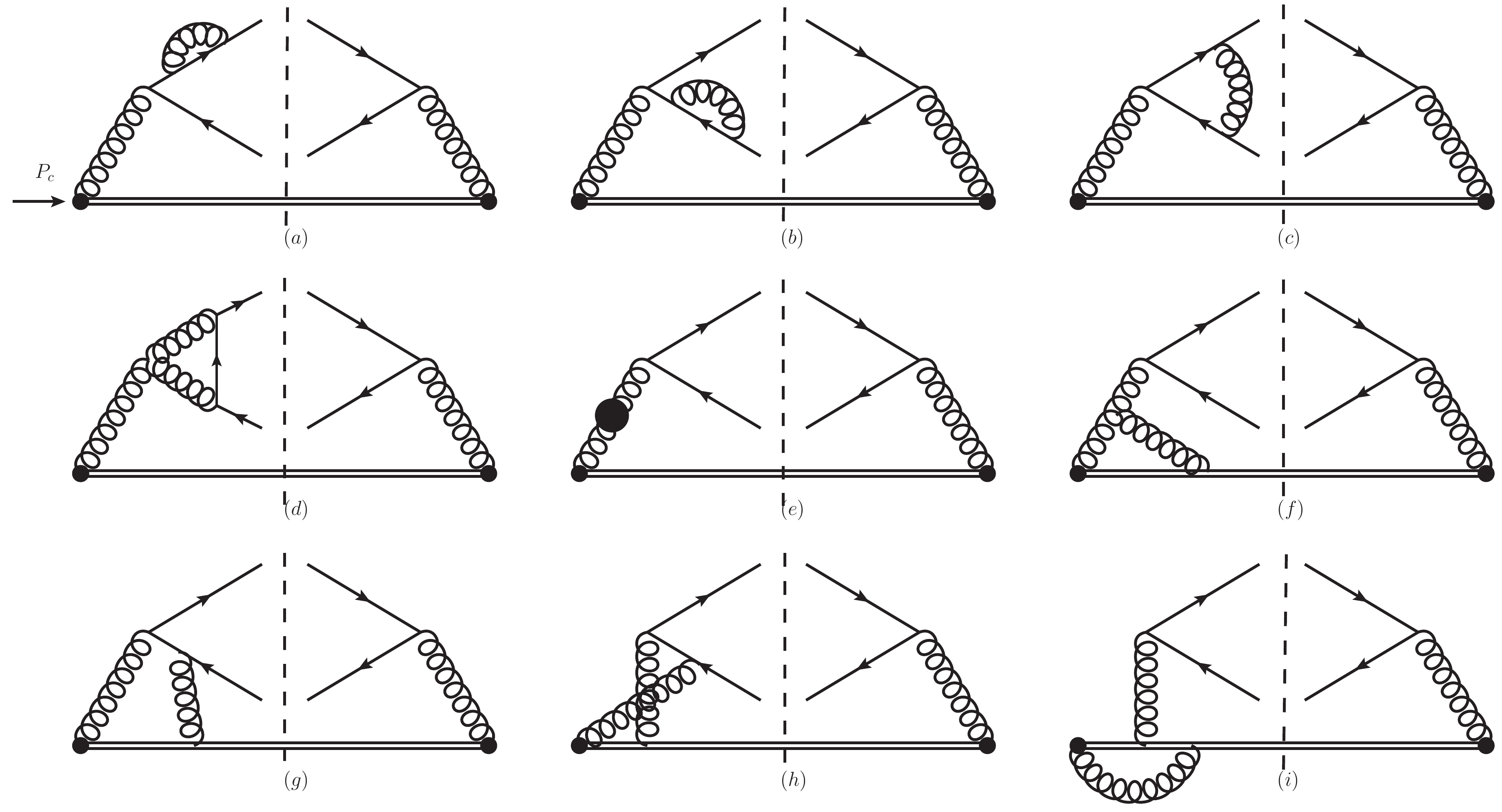}
		\caption{Feynman diagrams for virtual correction of the FF at NLO in Feynman gauge. The double solid line represents the gauge link along $l$ direction. The
			complex conjugate diagrams are not shown, but are included in the calculations.
			\label{fig:FFVirtualDiagram}}
	\end{center}
\end{figure}

At the NLO level, we only consider the FF $D_{g\to Q\bar Q [SS] }$, which consists
of virtual corrections and real corrections. The diagrams with virtual-gluon
corrections are shown in Fig.~\ref{fig:FFVirtualDiagram}. To compute the contributions
of these diagrams, we rewrite the integral of solid angle as
\begin{align}
\int \frac{d^{d-2}\Omega}{N_\Omega} =  \frac{2M_H}{|\boldsymbol{q}|^{d-3}}\int \frac{d^d
q}{N_\Omega}\delta(q^2+|\boldsymbol{q}|^2) \delta(P_H\cdot q).
\end{align}
We then change all delta functions to propagator denominators~\cite{Anastasiou:2002yz} by using
\begin{align}
\delta(x)=\frac{i}{2\pi} \lim_{\varepsilon \to 0} \biggr(\frac{1}{x+i\varepsilon}-\frac{1}{x-i\varepsilon}\biggr).
\end{align}
Then the obtained new Feynman integrals can be calculated analytically by using established multi-loop calculation techniques, like the IBP reduction~\cite{Chetyrkin:1981qh,Smirnov:2012gma,Smirnov:2014hma,Smirnov:2019qkx,Lee:2013mka} and  differential equations~\cite{Kotikov:1990kg,Remiddi:1997ny,Argeri:2007up,Henn:2013pwa,Henn:2014qga,Lee:2014ioa},
which results in
\begin{subequations}\label{eq:vir}
\begin{align}
D^{NLO}_{g\to Q\bar Q [SS]} (z ,M_H,m_Q,\mu) \biggr\vert_{vir.}^{a+b} =&  \bar D^{LO}(M_H,m_Q, \mu)\delta(1-z)\frac{\alpha_s}{2\pi}\frac{N_c^2-1}{2N_c} \biggr( \frac{4\pi\mu^2e^{-\gamma_E}}{m_Q^2}
\biggr)^{\epsilon}
\nonumber\\
& \times\biggr[ -\frac{3}{\epsilon}-4\biggr],
\\
D^{NLO}_{g\to Q\bar Q [SS]} (z,M_H,m_Q,\mu) \biggr\vert_{vir.}^{c} =&  \bar D^{LO}(M_H,m_Q, \mu)\delta(1-z) \frac{\alpha_s}{2\pi} \biggr( -\frac{1}{2N_c} \biggr) \biggr(\frac{
4\pi\mu^2e^{-\gamma_E}}{m_Q^2} \biggr)^{\epsilon}
\nonumber\\
& \times \biggr[ \frac{1}{\epsilon} \biggr( 1 + \frac{1+\Delta^2}{\Delta}\ln\frac{1+\Delta}{1- \Delta} \biggr) +  \mathcal {A}( \Delta)
\biggr],
\\
D^{NLO}_{g\to Q\bar Q [SS] } (z,M_H,m_Q,\mu)\biggr\vert_{vir.}^{d} =&  \bar D^{LO}(M_H,m_Q, \mu)\delta(1-z) \frac{\alpha_s}{2\pi}  \frac{N_c}{2} \biggr(\frac{ 4 \pi\mu^2e^{-\gamma_E}}{m_Q^2}
\biggr)^{\epsilon}
\nonumber\\
& \times \biggr[ \frac{3}{\epsilon} + \mathcal {B}(\Delta) \biggr],
\\
D^{NLO}_{g\to Q\bar Q [SS]} (z,M_H,m_Q,\mu)\biggr\vert_{vir.}^{e} =&  \bar D^{LO}(M_H,m_Q, \mu)\delta(1-z) \frac{\alpha_s}{2\pi}  \biggr(\frac{ 4 \pi\mu^2e^{-\gamma_E} }{M_H^2} \biggr)^{\epsilon}
\nonumber\\
& \times
\biggr[ \frac{5N_c-2n_f}{3 }\frac{1}{\epsilon} +\frac{31N_c-10n_f}{9 } \biggr],
\\
D^{NLO}_{g\to Q\bar Q [SS]} (z,M_H,m_Q, \mu) \biggr \vert_{vir.}^{f} =&  \bar D^{LO}(M_H,m_Q, \mu)\delta(1-z) \frac{\alpha_s}{2\pi}   N_c  \biggr(\frac{4\pi\mu^2e^{-\gamma_E}}{M_H^2}
\biggr)^{\epsilon}
\nonumber\\
& \times
\biggr[ \frac{2}{\epsilon^2}+ \frac{1}{\epsilon}  + 2 -\frac{7\pi^2}{6}  \biggr],
\\
D^{NLO}_{g\to Q\bar Q [SS]} (z ,M_H,m_Q,\mu) \biggr \vert_{vir.}^{g+h} =&  \bar D^{LO}(M_H,m_Q, \mu)\delta(1-z) \frac{\alpha_s}{2\pi}    N_c
\biggr(\frac{4\pi\mu^2e^{-\gamma_E}}{M_H^2} \biggr)^{\epsilon}
 \nonumber\\
&\hspace{-2cm} \times
\biggr[ -\frac{1}{\epsilon^2} -
2\biggr(1-\frac{1}{2\Delta} \ln \frac{1+\Delta}{1-\Delta}
 +\frac{M_H^2}{12(M_H+2m_Q)(M_H+m_Q)}
 \nonumber\\
&\hspace{-2cm} \times \biggr( 3- 2\Delta^2 -  \frac{3(1-\Delta^2)}{2\Delta} \ln \frac{1+\Delta}{1-\Delta} \biggr)\biggr)\frac{1}{\epsilon} + \mathcal {C}( \Delta) \biggr].
\\
D^{NLO}_{g\to Q\bar Q [SS]} (z ,M_H,m_Q,\mu) \biggr \vert_{vir.}^{i} =& 0.
\end{align}
\end{subequations}
In the above results we have dropped imaginary parts that are irrelevant for our purpose. The functions $\mathcal {A}(\Delta), \mathcal {B}(\Delta)$ and
$\mathcal {C}(\Delta)$ are infrared finite, but $\mathcal {A}(\Delta)$ contains Coulomb divergence.

\begin{figure}[htb!]
	\begin{center}
		\includegraphics[width=1\textwidth]{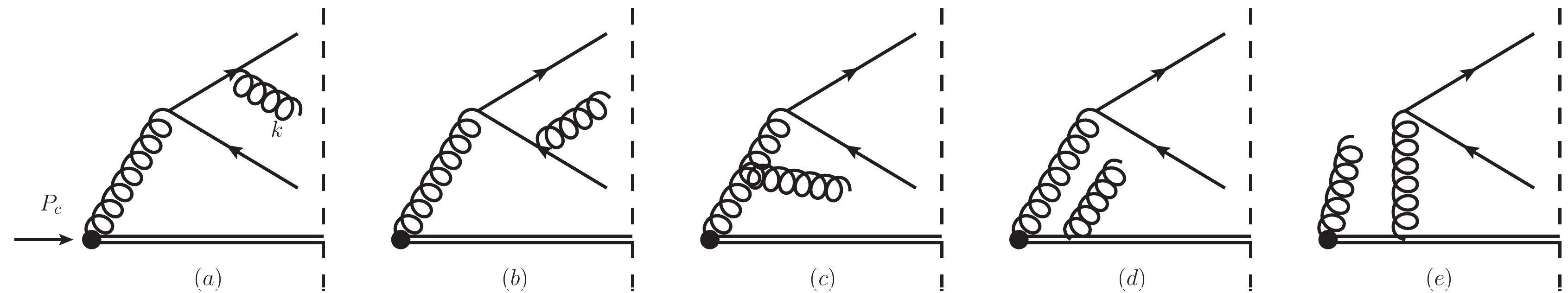}
		\caption{Feynman diagrams for real correction of the fragmentation function at NLO.
			\label{fig:FFrealDiagram}}
	\end{center}
\end{figure}

Feynman diagrams for the real corrections to $ D^{NLO}_{g\to Q\bar Q [SS]} $ are shown in Fig.~\ref{fig:FFrealDiagram}, where only diagrams on the left-hand side of the cut are shown.
For real corrections, the final state phase space in Eq.~\eqref{eq:FF0} reads
\begin{align}\label{eq:phase-1}
  d \Phi^{NLO} &= 	
    \frac{d^{d} P_{c}}{(2\pi)^{d}}
    \frac{d^{d} k}{(2\pi)^{d}}
    (2\pi)^{d} \delta^{d} \left(
    P_{c} - P_H -k \right) \delta \left( P_c^+ - \frac{P_H^{+}}{z} \right) 2\pi \delta(k^2)\theta(k^+),
\end{align}
where $k$ is the momentum of the final-state gluon. In the calculation, we first integrate out the solid angle. To do this, we decompose the amplitude $\mathcal{M}_{L}^{\rho}$ into the following form
\begin{align}
\mathcal{M}_{L}^{\rho}(P_H,P_c,m_Q,k,\lambda)=&
M_{L0}^{\rho} \int\frac{d^{d-2}\Omega}{N_\Omega} \frac{1}{(P_H+2q)\cdot k} +
M_{L0}^{\prime\rho} \int\frac{d^{d-2}\Omega}{N_\Omega}\frac{1}{(P_H-2q)\cdot k}
\nonumber\\
& \hspace{-2cm} +M_{L1}^{\rho\mu} \int\frac{d^{d-2}\Omega}{N_\Omega}\frac{q_\mu}{(P_H+2q)\cdot k}
+M_{L1}^{\prime\rho\mu} \int\frac{d^{d-2}\Omega}{N_\Omega}\frac{q_\mu}{(P_H-2q)\cdot k}
\nonumber\\
& \hspace{-2cm}+M_{L2}^{\rho\mu\nu} \int\frac{d^{d-2}\Omega}{N_\Omega}\frac{q_\mu q_\nu}{(P_H+2q)\cdot k}+M_{L2}^{\prime\rho\mu\nu} \int\frac{d^{d-2}\Omega}{N_\Omega}\frac{q_\mu q_\nu}{(P_H-2q)\cdot k}\nonumber\\
& \hspace{-2cm} +M_{L3}^{\rho\mu\nu\tau} \int\frac{d^{d-2}\Omega}{N_\Omega}\frac{q_\mu q_\nu q_\tau}{(P_H+2q)\cdot k}+M_{L3}^{\prime\rho\mu\nu\tau} \int\frac{d^{d-2}\Omega}{N_\Omega}\frac{q_\mu q_\nu q_\tau}{(P_H-2q)\cdot k},
\end{align}
where the functions $M_{Li}$ and $M_{Li}^{\prime}$ are $q$ independent. Then the integral over $\Omega$ can be performed by using equations provided in Appendix.~\ref{sec:app-angle}. After that, we carry out the integration over the phase space $d \Phi^{(1)}$, which results in
\begin{align}\label{eq:real}
D^{NLO}_{g \to Q\bar Q [SS] }(z,M_H,m_Q, \mu)\biggr\vert_{real} =&
\bar D^{LO}(M_H,m_Q, \mu)\biggr(\frac{4\pi \mu^2 e^{-\gamma_E}}{ M_H
^2}\biggr)^{\epsilon} \frac{\alpha_s  N_c}{\pi} \nonumber\\
 &\hspace{-4.5cm} \times \biggr[ - \frac{\delta(1-z)}{2 \epsilon^2}
  -  \frac{1}{2 \epsilon} \frac{4\delta(1-z)}{3(M_H + 2m_Q)^2(M_H+m_Q)^2}  \biggr( m_Q^4( 9\mathcal {T}^2 - 9\mathcal {T} - 1)
\nonumber\\
 &\hspace{-4.5cm}   + \frac{3}{2}m_Q^3 M_H(3\mathcal {T} -8)
   - 10m_Q^2 M_H^2 -\frac{39}{8}m_Q M_H^3 - \frac{13}{16}M_H^4 \biggr)
 \nonumber\\
 &\hspace{-4.5cm}
 + \frac{1}{\epsilon} \frac{1}{(1-z)_+} \frac{(z^2-z+1)^2}{z}
  + \mathcal
    {R}(z,M_H,\Delta)  \biggr] ,
\end{align}
where $\mathcal{R}(z,M_H,\Delta)$ is infrared finite.

Combining the virtual and real corrections in Eq.~\eqref{eq:vir} and Eq.~\eqref{eq:real}, we obtain NLO
correction for the FF $D_{g \to Q\bar Q[SS]}$, which  contains
ultraviolet and infrared divergences in the form of single poles in $\epsilon$ as shown below
\begin{align}\label{eq:NLOFF-div}
 D^{NLO}_{g \to Q\bar Q [SS] }(z,M_H,m_Q, \mu)\biggr\vert_{div.} =&
\bar D^{LO}(M_H,m_Q, \mu)
\frac{\alpha_s  }{\pi} (4\pi \mu^2 e^{-\gamma_E} )^{\epsilon}
\nonumber\\ &\hspace{-5cm} \times \biggr\{ \frac{1}{\epsilon} \biggr[ \beta_0 \delta(1-z) + \frac{N_c}{(1-z)_+}\frac{(z^2-z+1)^2}{z} \biggr]
 - \frac{1}{\epsilon}\delta(1-z)
  N_c
\nonumber  \\
 &\hspace{-5cm}\times
 \biggr[1-\frac{1}{2\Delta} \ln \frac{1+\Delta}{1-\Delta}
 +\frac{M_H^2}{12(2m_Q^2 +3m_QM_H+M_H^2)}
 \biggr( 3- 2\Delta^2 -  \frac{3(1-\Delta^2)}{2\Delta} \ln \frac{1+\Delta}{1-\Delta} \biggr) \biggr]
\nonumber \\
 &\hspace{-5cm}
 -  \frac{1}{ \epsilon} \frac{2N_c \delta(1-z)}{3(M_H + 2m_Q)^2(M_H+m_Q)^2} \biggr[ m_Q^4( 9\mathcal {T}^2 - 9\mathcal {T} - 1)
   + \frac{3}{2}m_Q^3 M_H(3\mathcal {T} -8)
   - 10m_Q^2 M_H^2
\nonumber \\
&\hspace{-5cm}
   -\frac{39}{8}m_Q M_H^3
  - \frac{13}{16}M_H^4 \biggr]
     - \delta(1-z) \biggr[ \frac{N_c}{2}  + \frac{1}{2N_c} \biggr(  \frac{1+\Delta^2}{2\Delta}\ln\frac{1+\Delta}{1-
    \Delta} -1 \biggr) \biggr] \frac{1}{\epsilon} \biggr\}.
\end{align}
Ultraviolet divergences in the above equation are canceled by the renormalization of the coupling constant
$\alpha_s$ and the operator defining the FF.
The renormalization of $\alpha_s$ in the $\overline{\textrm{MS}}$ scheme
can be carried out by making the following replacement in Eq.~\eqref{eq:FFLO}
\begin{equation}
\alpha_s \rightarrow \alpha_s\biggr( 1 - \frac{\alpha_s}{2\pi} (4\pi e^{-\gamma_E})^\epsilon \beta_0
\frac{1}{\epsilon} \biggr),
\end{equation}
and the operator renormalization in the $\overline{\textrm{MS}}$ scheme can be carried out by further making the following
replacement
\begin{align}
D^{LO}_{g\to Q\bar Q [SS]} (z,M_H,m_Q, \mu) \to& D^{LO}_{g\to Q\bar Q [SS]}
(z,M_H,m_Q, \mu) - \frac{\alpha_s}{2\pi} (4\pi e^{-\gamma_E})^\epsilon \frac{1}{\epsilon}\nonumber\\
&\times
\int_z^1 \frac{dx}{x} P_{gg}^{LO}(x) D^{LO}_{g\to Q\bar Q [SS]} (z/x,M_H,m_Q, \mu) .
\end{align}
Then the renormalized $D_{g \to Q\bar Q[SS]}$ reads
\begin{align}\label{eq:NLOFF-R}
 D^{NLO}_{g \to Q\bar Q [SS] }(z,M_H,m_Q, \mu) =&
\bar D^{LO}(M_H,m_Q, \mu)
\frac{\alpha_s  }{\pi} (4\pi \mu_c^2 e^{-\gamma_E} )^{\epsilon}
 \biggr\{
 - \frac{1}{\epsilon_{IR}}N_c\delta(1-z)
\nonumber  \\
 &\hspace{-4.5cm}\times
 \biggr[1-\frac{1}{2\Delta} \ln \frac{1+\Delta}{1-\Delta}
 +\frac{M_H^2}{12(2m_Q^2 +3m_QM_H+M_H^2)}
 \biggr( 3- 2\Delta^2 -  \frac{3(1-\Delta^2)}{2\Delta} \ln \frac{1+\Delta}{1-\Delta} \biggr) \biggr]
\nonumber \\
 &\hspace{-4.5cm}
 -  \frac{1}{ \epsilon_{IR}} \frac{2N_c \delta(1-z)}{3(M_H + 2m_Q)^2(M_H+m_Q)^2} \biggr[ m_Q^4( 9\mathcal {T}^2 - 9\mathcal {T} - 1)
   + \frac{3}{2}m_Q^3 M_H(3\mathcal {T} -8)
\nonumber \\
&\hspace{-4.5cm}
   - 10m_Q^2 M_H^2
   -\frac{39}{8}m_Q M_H^3
  - \frac{13}{16}M_H^4 \biggr]
     - \frac{1}{\epsilon_{IR}}\biggr[ \frac{N_c}{2}  + \frac{1}{2N_c} \biggr(  \frac{1+\Delta^2}{2\Delta}\ln\frac{1+\Delta}{1-
    \Delta} -1 \biggr) \biggr]
\nonumber \\
&\hspace{-4.5cm} \times \delta(1-z)  + \mathcal
    {F}(z,M_H, \Delta,\mu) \biggr\}.
\end{align}
Here we have distinguished between the dimensional regularization scale $\mu_c$ and  the collinear factorization scale $\mu$. All remaining divergences in the above equation are infrared divergences, and they should be absorbed into the definition of SGDs if the SGF is valid at NLO. The infrared finite functions $\mathcal {A}, \mathcal {B}, \mathcal {C}, \mathcal {R}, \mathcal {F}$ in above equations are listed in App.~\ref{sec:app-finite-function}. They also can be read from the ancillary file.

\subsection{Matching the short-distance hard part}\label{sec:FF-matching}

Inserting Eqs.~\eqref{eq:FFLO}, \eqref{eq:NLOFF-R}, \eqref{eq:SGD-LO1}, \eqref{eq:SGD-LO2}, \eqref{eq:SGD-R} and~\eqref{eq:SGDs} into
the matching equation Eq.~\eqref{eq:matchingEq3}, we obtain the hard part $\hat{D}_{[SS]}$ up to NLO.  The result is listed in App.~\ref{sec:app-finite-function} and we also provide it in the ancillary file. In the result we have chose $\mu_f=\mu$. We find that all infrared and Coulomb divergences in Eq.~\eqref{eq:NLOFF-R} can be correctly subtracted by the SGDs. Thus we conclude that the SGF formula Eq.~\eqref{eq:FFSGF-1} holds at least at the one-loop level.

Based on $\hat{D}_{[SS]}$,
we can easily obtain the short distance
hard part at leading order
in the velocity expansion, which reads
\begin{subequations}\label{eq:SGF-SDCNLO}
\begin{align}
&\hat{D}_{[SS]}^{LO,(0)}(\hat z,M_H/x,  \mu, \mu_f) =  \frac{\pi \alpha_s}{(N_c^2-1)}\frac{8x^3}{M_H^3}  \delta(1-\hat z),\label{eq:SGF-SDCNLOa}\\
& \hat{D}_{[SS]}^{NLO,(0)}(\hat{z},M_H/x, \mu, \mu_f)
\nonumber\\
&= \frac{4\alpha_s^2N_c x^3}{(N_c^2-1)M_H^3}  \biggr[ \frac{1}{2}\delta(1-\hat{z})  \biggr(2A(\mu,M_H/x)  + \frac{2\beta_0}{N_c}\ln\biggr(\frac{x^2\mu_f^2 e^{ - 1}}{M_H^2}\biggr) + \ln^2\biggr(\frac{x^2\mu_f^2 e^{ - 1}}{M_H^2}\biggr)
\nonumber\\
& \quad
+\frac{\pi^2}{6} -1 \biggr)
+  \frac{1}{N_c} P^{(0)}_{gg}(\hat{z}) \ln \biggr( \frac {\mu^2}{\mu_f^2} \biggr)
+  \biggr( \frac{2(1-\hat{z})}{\hat{z}} + \hat{z}( 4+2\hat{z}^2)+ \frac{2\hat{z}^4}{9}(5+\hat{z}) \biggr)
\nonumber\\
&\quad \times \biggr(\ln\biggr( \frac{x^2\mu_f^2 e^{-1}}{M_H^2}\biggr)-2 \ln(1-\hat{z})\biggr)
  + \frac{2(1-\hat{z})}{\hat{z}}
-  \biggr( \frac{4\hat{z}^4}{1-\hat{z}}- \frac{4\hat{z}^4}{9}(5+\hat{z})\biggr) \ln \hat{z} \biggr] .
\end{align}
\end{subequations}
The factorization scale $\mu_f$ should be chosen at the order of $M_H$ in order to avoid
the appearance of large logarithms of $M_H/\mu_f$.
Besides, we find the hard part is also free of the threshold logarithms in the $\hat z \rightarrow 1$ limit with the scale choice $\mu_f=\mu=M_H$. The dependence of $\mu$ and $\mu_f$ in hard part can be recovered by RGEs obeyed by the gluon FF and the color-octet $^3S_1$ SGD.

We eventually obtain gluon FFs in $\COcSa$ channel
\begin{subequations}\label{eq:FFSGF-res}
	\begin{align}
	D_{g \to H}(z,M_H,m_Q,M_H) = & \int_{z}^1 \frac{dx}{x}   \hat{D}_{[SS]}(\hat{z},M_H/x,m_Q, M_H,M_H)
	\nonumber\\&\times
	F_{[SS] \to H}(x,M_H,m_Q,M_H), \\
	D_{g \to H}^{(0)}(z,M_H,m_Q,M_H) = & \int_{z}^1 \frac{dx}{x}   \hat{D}_{[SS]}^{(0)}(\hat{z},M_H/x, M_H,M_H)
	\nonumber\\&\times
	F_{[SS] \to H}(x,M_H,m_Q,M_H).
	\end{align}
\end{subequations}
Using the DGLAP equation~\eqref{eq:DGLAP} and initial
condition Eq.~\eqref{eq:FFSGF-res}, we can obtain the gluon FFs at any larger scale.

\subsection{Numerical results}\label{sec:Numerical-results}

We now present our numerical results for the gluon FFs given in Eq.~\eqref{eq:FFSGF-res}. We use the model function Eq.~\eqref{eq:model-SGD} with overall normalization $N_H=\langle O^H(\state{{3}}{S}{1}{8}) \rangle/3=1/3$ and $b=2$. We choose $m_Q=1.4 \mathrm{GeV}$, $M_H=3.1 \mathrm{GeV}$,  $\Lambda_{QCD}^{(4)}= 0.217 \mathrm{GeV} $
($\Lambda_{QCD}^{(4)}= 0.338 \mathrm{GeV} $ ) for LO (NLO), and $n_f = 3$ in $\beta_0$ which means contributions from virtual or initial heavy quarks are ignored. In the following, in not specified we refer to fully NLO FF $D_{g \to H}(z,M_H,m_Q,M_H)$ calculated in SGF with $\bar{\Lambda}=0.6\mathrm{GeV}$. We note that, as NLO evolution kernel for SGD is still not available, all results presented in this paper are obtained by using the LO evolution kernel~\eqref{eq:SGD-kernel}.

\begin{figure}[htb!]
 \begin{center}
 \vspace*{0.8cm}
 \hspace*{-5mm}
 \includegraphics[width=0.45\textwidth]{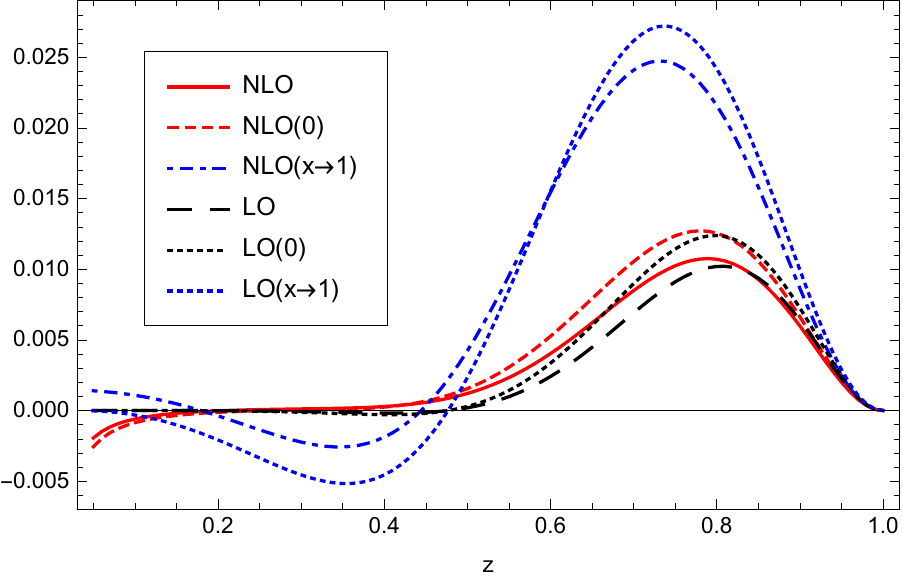}
 \hspace*{5mm}
 \includegraphics[width=0.45\textwidth]{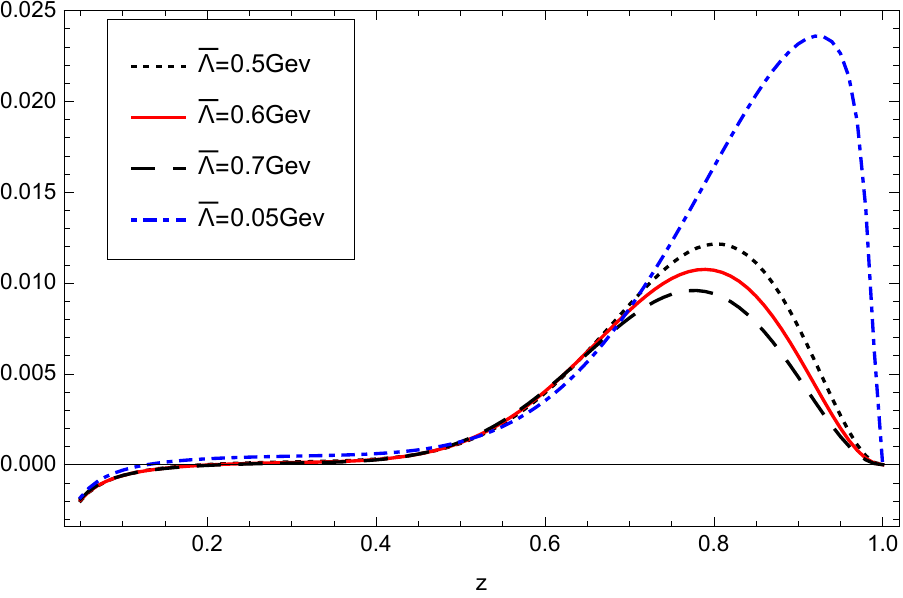}
 \end{center}
 \vspace*{-.5cm}
 \caption{Left figure: Comparison of the gluon FF obtained in different approximations.   Right figure: $\bar{\Lambda}$ dependence of gluon FF at NLO. \label{fig:FFmu0}}
 \vspace*{0.cm}
\end{figure}

In the left figure of Fig.~\ref{fig:FFmu0} we show the gluon FF obtained by different approximations, where $\text{LO}{(0)}$ and $\text{NLO}{(0)}$ represent the lowest order in $v^2$ expansion $D_{g \to H}^{(0)}$ at the corresponding $\alpha_s$ order. We find that  $D^{(0)}_{g \to H}$  is a good approximation to $D_{g \to H}$, with deviation smaller than $20\%$. This implies that the convergence of velocity expansion in SGF is good and that $D_{g \to H}$ is insensitive to the input of heavy quark mass $m_Q$. For comparison, we also provide plots of FFs obtained by setting $x=1$ in hard part \eqref{eq:SGF-SDCNLO} everywhere except in the delta function $\delta(1-{z}/x)$. This is clear a bad approximation because it overshoots original results by at least a factor of 2, which indicates big corrections at high order in $1-x$ expansion. \footnote{These are also velocity corrections when $1-x\approx v$ or $v^2$. Resumming this series of velocity corrections is the main motivation of SGF.} This can be easily understood because the hard part are proportional to $x^3$ and SGD is peaked around $x=0.75$. We note that, if one uses the shape function method \cite{Fleming:2003gt,Fleming:2006cd,Bauer:2001rh,Beneke:1997qw} to calculate the gluon FF, one should take $x\to 1$ in hard part, which will result in large relativistic corrections as discussed above.

In the right figure of Fig.~\ref{fig:FFmu0} we show the $\bar{\Lambda}$ dependence by choosing it to $0.05\mathrm{GeV}$, $0.5\mathrm{GeV}$, $0.6\mathrm{GeV}$ and $0.7\mathrm{GeV}$. We find that the gluon FF at small $z$ is much less affected by the parameter $\bar{\Lambda}$ than those at large $z$, which indicates that gluon FF at small $z$ is dominated by perturbative effects while that at large $z$ is sensitive to nonperturbative dynamics. This sensitivity provides a possibility to extract  $\bar{\Lambda}$  using experimental data.

\begin{table}
	\caption[]{The ratios of moments of fragmentation functions in different approaches defined in Eq.~\eqref{eq:ratio}. $R^{(0)}$ represents FF with lowest order in $v^2$ expansion, $R^{0.05}$ represents FF with $\bar{\Lambda}=0.05\mathrm{GeV}$, $R^{0.05(\text{pert.})}$ represents perturbative expansion of FF with $\bar{\Lambda}=0.05\mathrm{GeV}$ and also replacing $M_H$ by $2m_Q$, and $R^{\text{NRQCD}}$ represents FF calculated in NRQCD.  \label{table:Moment}}
	\renewcommand{\arraystretch}{1.5}
	\[
	\begin{tabular}{|c|c|c|c|c|c|}
	\hline\hline
	\multirow{1}{*}{Factor} & \multirow{1}{*}{$z^2$}& \multirow{1}{*}{$z^3$} & \multirow{1}{*}{$z^4$}& \multirow{1}{*}{$z^5$} & \multirow{1}{*}{$z^6$} \\
	\cline{1-6}
	 $R^{(0)}$ & 1.18& 1.18& 1.17& 1.16& 1.16 \\
	\hline
	$R^{0.05}$ & 2.68& 2.91& 3.16& 3.42& 3.69 \\
	\hline
	$R^{0.05 (\text{pert.})}$ & 7.81 & 7.71 & 7.36 & 6.70 & 5.63  \\
	\hline
	$R^{\text{NRQCD}}$ & 7.54 & 7.48 & 7.16 & 6.49 & 5.34  \\
	\hline
	\hline
	\end{tabular}
	\]
	\renewcommand{\arraystretch}{1.0}
\end{table}

We note that the plot with $\bar{\Lambda}=0.05\mathrm{GeV}$ in Fig.~\ref{fig:FFmu0} is to mimic the case with $\bar{\Lambda}\to 0$. Although nonperturbative input models with  $\bar{\Lambda}=0.6 \mathrm{GeV}$ and $0.05\mathrm{GeV}$ have normalized to the same value, they result in significantly different FFs, mainly due to the large logarithm resummation. To quantify the difference, we calculate the $n$-th moment of FFs and define the following ratios,
\begin{align}\label{eq:ratio}
R^{X}(n) \equiv & \frac{\int_0^1 dz z^n D^{X}_{g \to H}(z,M_H,m_Q,\mu)}{\int_0^1 dz z^{n} D_{g \to H}(z,M_H,m_Q,\mu) },
\end{align}
with numerical results given in Table.~\ref{table:Moment} for $n=2,3,4,5,6$. 
By taking $D_{g \to H}$ as “exact” result, the values of $R$ indicate that the lowest order in velocity expansion in SGF (denoted as $R^{(0)}$) is a good approximation, while changing $\bar{\Lambda}$ from $0.6 \mathrm{GeV}$ to $0.05\mathrm{GeV}$  (denoted as $R^{0.05}$) results in large deviation, about a factor of 3.

\begin{figure}[htb!]
	\begin{center}
		\vspace*{0.8cm}
		\hspace*{-5mm}
		\includegraphics[width=0.45\textwidth]{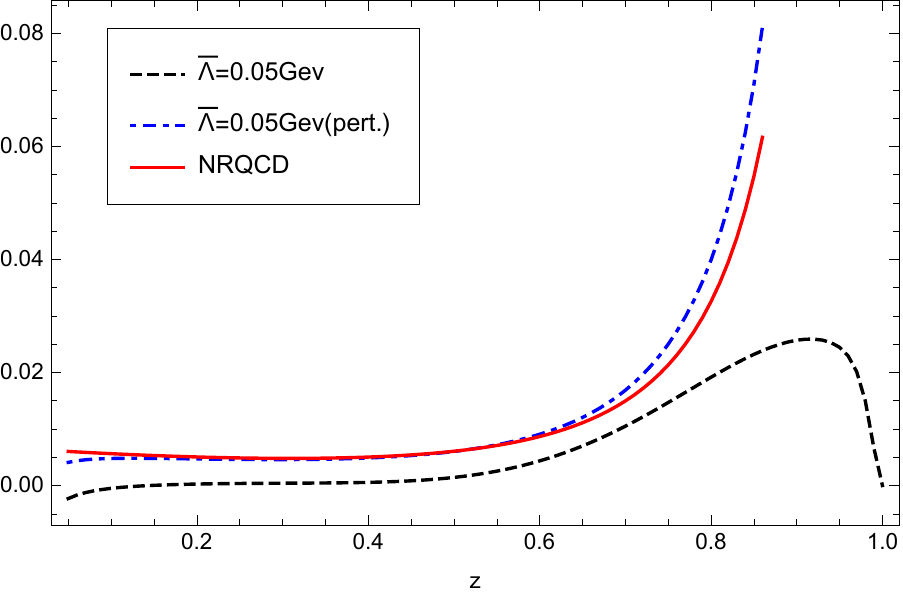}
	\end{center}
	\vspace*{-.5cm}
	\caption{Comparison of the gluon FFs obtained in SGF and NRQCD, where ``pert.'' means expanding $D^{\bar{\Lambda}=0.05 \mathrm{GeV}}_{g \to H}$ to NLO in $\alpha_s$ and replacing $M_H$ by $2m_Q$. \label{fig:NRQCD}}
	\vspace*{0.cm}
\end{figure}

As shown in Fig.~\ref{fig:NRQCD}, by perturbatively expanding $D^{\bar{\Lambda}=0.05 \mathrm{GeV}}_{g \to H}$ to NLO in $\alpha_s$ and replacing $M_H$ by $2m_Q$ (denoted as ``pert.''), we can nicely recover NRQCD result~\cite{Ma:1995ci,Braaten:2000pc,Ma:2013yla}. Their deviations at end points of $z$ are due to the nonzero $\bar{\Lambda}$, but they have almost the same moments as shown in Table.~\ref{table:Moment}, where $R^{\text{NRQCD}}$ means that the numerator of the ratio is the NRQCD FFs. We thus find that the NRQCD results overshoot SGF results with our prefered treatment by a factor about $5-7$, which is consistent with the conclusion in Ref.~\cite{Ma:2017xno}. One of the main reasons for deviation is due to the large logarithm resummation, and the other one is due to the choice of $\bar{\Lambda}=0.6 \mathrm{GeV}$ instead of zero.

\section{Summary}\label{sec:Conclusion}

In this paper we studied the FF from the gluon to color-octet $^3S_1$ heavy quark-antiquark pair in SGF approach, which is expressed as
the convolution of perturbative short-distance hard part with the one-dimensional color-octet $^3S_1$ SGD in Eq.~\eqref{eq:general-FFSGF-cut}.  We calculated the color-octet $^3S_1$ SGD to NLO in Eq.~\eqref{eq:SGD-R} and derived a RGE for it in Eq.~\eqref{eq:general-RGE-SGDs}. By solving the RGE, we resummed the threshold logarithms to all order in perturbation theory.
Based on this, we perturbatively calculated the short distance hard part of the gluon FF up to NLO in Eq.~\eqref{eq:NLOhard}. 
With a natural scales choice, the hard part is free of threshold logarithms, which has been resummed by the RGE of SGD. This demonstrated the validity of SGF at NLO.

Our numerical results shown that velocity expansion in the SGF has a good convergence, and the lowest order approximation can
already capture most physics. By a specific choice of parameters (but not prefered), the SGF can reproduce the NRQCD results. But then we get a big deviation, as large as a factor of 6, from our prefered results. This implies that velocity expansion in NRQCD has a very slow convergence. We also show that even the large threshold logarithms in NRQCD results are resumed by using shape function method, there should be still significant velocity corrections.

Our results may provide a new insight to understand
the mechanisms of quarkonium production, especially for quarkonium production at high energy colliders and quarkonium production within a jet~\cite{Baumgart:2014upa,Kang:2017yde}.

\begin{acknowledgments}

We thank Kuang-Ta Chao, Xiaohui Liu, Shuai Zhao, Xu-Chang Zheng and Peng Zhang for many useful discussions. The work is supported in part by the National Natural Science Foundation of China (Grants No. 11875071, No. 11975029) and the China Postdoctoral Science Foundation under Grant No.2018M631234.

\end{acknowledgments}

\appendix
\renewcommand{\theequation}{\thesection\arabic{equation}}

\section{Solving RGE of SGD} \label{sec:app-rge}

Here we give the details of solving the RGE in Eq.~\eqref{eq:LaplaceRGE}. Integrating Eq.~\eqref{eq:LaplaceRGE} from a initial scale $\mu_0$ to $\mu_f$ by changing
variables to $\alpha_s$ with $d \ln\mu= d \alpha_s/\beta(\alpha_s)$ gives the solution
\begin{align}\label{eq:SGDsolution}
\tilde F_{[SS] \to H}(s , M_H,m_Q, \mu_f) =&  \mathrm{exp}\biggr[ \eta(\mu_f, \mu_0)  \ln  ( \bar s \mu_0   )
\biggr]e^{V(\mu_f,\mu_0)}
\nonumber\\
&  \times
\tilde F_{[SS] \to H}(s , M_H,m_Q, \mu_0),
\end{align}
the evolution functions are given by
\begin{align}\label{eq:evolution-function}
\eta(\mu_f, \mu_0) =& - \int_{\alpha_s(\mu_0)}^{\alpha_s(\mu_f)} \frac{d\alpha}{\beta(\alpha)} \frac{\alpha_s \Gamma^F_{0}}{4\pi} , \nonumber\\
V(\mu_f, \mu_0) =& - \int_{\alpha_s(\mu_0)}^{\alpha_s(\mu_f)} \frac{d\alpha}{\beta(\alpha)} \frac{\alpha_s \gamma^F_0}{4\pi}- \int_{\alpha_s(\mu_0)}^{\alpha_s(\mu_f)} \frac{d\alpha}{\beta(\alpha)} \frac{\alpha_s \Gamma^F_{0}}{4\pi} \int_{\alpha_s(\mu_0)}^{\alpha}\frac{d\alpha'}{\beta(\alpha')}.
\end{align}
Defining $r = 1 -\alpha_s(\mu_f)/\alpha_s(\mu_0)$ and inserting $\beta(\alpha_s) = -\beta_0 \alpha_s^2 /\pi + \mathcal {O}(\alpha_s^3) $ into Eq.~\eqref{eq:evolution-function}, we obtain following expression for the evolution functions
\begin{align}\label{eq:RGE-function}
\eta(\mu_f, \mu_0) =& \frac{\Gamma^F_0}{4\beta_0}  \ln(1-r), \nonumber\\
V(\mu_f, \mu_0) =& \frac{ \Gamma^F_0}{(4\beta_0)^2} \biggr[ - \frac{4 \pi}{\alpha_s(\mu)}\biggr(r +(1-r)\ln(1-r) \biggr)  \biggr]+ \frac{ \gamma_0^F}{4\beta_0} \ln(1-r) .
\end{align}
With the help of the relation Eq.~\eqref{eq:SGDsolution}, the color-octet $^3S_1$ SGD can be evolved to the scale $\mu_f$ from a reference scale $\mu_0$, at which a nonperturbative model for the SGD is provided. And the exponential in Eq.~\eqref{eq:SGDsolution} resums the single and double logarithms of $\mu_f/\mu_0$.

To determine the initial scale $\mu_0$, we expand the $m_Q^2$ in perturbative SGD $F_{[SS] \to Q\bar Q[SS]}$ around $M_H^2/4$, we have
\begin{align}\label{eq:SGDQQ}
F_{[SS] \to Q\bar Q[SS]}(\omega, M_H,m_Q ,\mu_f) =& M_H \biggr\{ \delta(\omega) -\frac{\alpha_s N_c}{4\pi}\biggr[ \biggr( 4\ln^2 \mu_f  - 4\ln \mu_f +\frac{\pi^2}{6} + \frac{1}{N_c^2}\frac{\pi^2}{\Delta} \biggr)
\nonumber\\
& \times \delta(\omega) + \biggr[\frac{4}{\omega} \biggr]_+   + 8\biggr[\frac{\ln(\omega/\mu_f)}{\omega}\biggr]_+ \biggr] \biggr\} + \mathcal {O}(\alpha_s\Delta),
\end{align}
which result in
\begin{align}\label{eq:fixed-order}
\tilde {F}_{[SS] \to Q\bar Q[SS]}(s, M_H,m_Q ,\mu_0) =&
M_H  \biggr [ 1-\frac{\alpha_s N_c}{4\pi} \biggr( 4  \ln^2(\bar s \mu_0)- 4\ln(\bar s \mu_0 )
+ \frac{5\pi^2}{6}   + \frac{1}{N_c^2}\frac{\pi^2}{\Delta} \biggr)\biggr ]
\nonumber\\
&+ \mathcal {O}(\alpha_s\Delta).
\end{align}
We can find, in the Laplace space, the fixed-order
perturbative expansion of $\tilde{F}_{[SS] \to Q\bar Q[SS]}$ contains single and double logarithms of $ \bar s \mu_0 $
and these large logarithms can be minimized by the choice of scale
\begin{equation} \label{eq:scale-1}
\mu_0= 1/\bar {s}.
\end{equation}
Making the above characteristic scale choice in Eq.~\eqref{eq:SGDsolution}, we can resum those large logarithms of $s $ and obtain the resummed SGD, which reads
\begin{align}\label{eq:resummed-SGD}
\tilde F_{[SS] \to H}(s,M_H,m_Q,\mu_f=M_H) =& \tilde F_{[SS] \to H }(s, M_H,m_Q,  1/\bar s)
\nonumber\\
& \times \mathrm{exp}\biggr[ h_0(\chi) \ln (M_H \bar s)  + h_1(\chi) \biggr],
\end{align}
where
\begin{align}
\chi =& \frac{\beta_0}{2\pi}  \alpha_s(M_H) \ln (M_H \bar s).
\end{align}
The functions $h_{0,1}$ are given by
\begin{align}\label{eq:resummed-F}
h_0(\chi) =& \frac{\Gamma^F_0}{8\beta_0 \chi} \biggr( -2\chi - (1-2\chi) \ln(1-2\chi) \biggr), \nonumber\\
h_1(\chi) =&  \frac{\gamma^F_0}{4\beta_0}\ln(1-2\chi) .
\end{align}
Note that the term $\ln(1-2\chi)$ in above expressions gives rise to a cut singularity that start at the branch point
\begin{align}
\bar s_L  =& \frac{1}{M_H} \mathrm{exp}\biggr[ \frac{\pi}{\beta_0 \alpha_s(M_H)}\biggr]\simeq \frac{1}{\Lambda_{\mathrm{QCD}}}.
\end{align}
This singularity, as a result of the divergence of the running coupling $\alpha_s(\mu)$ near the Landau pole at $\mu=\Lambda_{\mathrm{QCD}}$, signals the onset of nonperturbative phenomena at very large $s$ or, equivalently, when $x$ is very close to the value $x =1$. To deal with the Landau singularity, we
introduce a cut-off $\bar {s}^\ast$ above which the functions $h_{0,1}$ remained constant, i.e. ``frozen'' at $\bar {s}^\ast$. Additionally, a nonperturbative model was introduced to correct the formalism for nonperturbative effects at very large $s$.
Following \cite{Cacciari:2005uk}, we make the replacement
\begin{align}\label{eq:s-ast}
\bar s \rightarrow  \bar s\frac{1+ 1/(M_H\bar {s}^\ast)}{ 1+  \bar s/ \bar {s}^\ast}, \quad \quad \bar {s}^\ast= \bar s_L/a
\end{align}
in Eq.~\eqref{eq:resummed-F}. Here $a$ is a parameter not smaller than one, but of order one. Such a replacement prevents $\chi$ from entering the nonperturbative regime. Then we model the initial SGD $\tilde F_{[SS] \to H}(s, M_H,m_Q,  1/\bar s)$ in following form
\begin{align}\label{eq:ModelSGD}
\tilde F_{[SS] \to H}(s, M_H,m_Q,  1/\bar s) =& \tilde F_{\textrm{fix}}(M_H)  \tilde F^{\textrm{mod}}(s) ,
\end{align}
with
\begin{subequations}\label{eq:fixorder}
	\begin{align}
	\tilde F_{\textrm{fix}}^{LO}(M_H)=&1, \\
	\tilde F_{\textrm{fix}}^{NLO}(M_H)=& - \frac{\alpha_s(M_H) N_c }{4\pi} \frac{5\pi^2}{6}.
	\end{align}
\end{subequations}
Here $\tilde F_{\textrm{fix}}(M_H)$ is introduced to recover the perturbative calculated $\tilde {F}_{[SS] \to Q\bar Q[SS]}$ with the Coulomb divergences and IR divergences (if they exist) be subtracted, as we have
\begin{align}
\tilde {F}_{[SS] \to Q\bar Q[SS]}(s, M_H,m_Q ,1/\bar s) =&
M_H  \biggr [ 1-\frac{\alpha_s N_c}{4\pi} \biggr(  \frac{5\pi^2}{6}   + \frac{1}{N_c^2}\frac{\pi^2}{\Delta} \biggr)\biggr ]
+ \mathcal {O}(\alpha_s\Delta).
\end{align}
And the model $\tilde F^{\textrm{mod}}(s)$ is introduced to describe the nonperturbative effects.

\section{Integration over solid angle}\label{sec:app-angle}
In this appendix, we give the expressions that integrate over the solid angle $\Omega$ of $q$ in the $Q \bar Q$ rest frame. For the momentum $k$ which satisfies $k^2=0$, we have following equations,
\begin{subequations}
\begin{align}
\int d^{d-2}\Omega \frac{1}{(P_H-2q)\cdot k} &=   \int d^{d-2} \Omega \frac{1}{(P_H+2q)\cdot k}
=\frac{\mathcal {T}}{P_H \cdot k} \int d^{d-2} \Omega ,\\
\int d^{d-2} \Omega \frac{q^\mu}{(P_H-2q)\cdot k} &=  \int d^{d-2} \Omega \frac{ - q^\mu}{(P_H+2q)\cdot k}
=\mathcal {F}_1 \mathbb{P}^{\mu \nu}k_\nu \int d^{d-2}\Omega , \\
\int d^{d-2}\Omega \frac{q^\mu q^\rho}{(P_H-2q)\cdot k} &=
\int d^{d-2}\Omega \frac{q^\mu q^\rho}{(P_H+2q)\cdot k}
\nonumber\\
&
= (\mathcal {F}_2 \mathbb{P}^{\mu \rho} + \mathcal {G}_2 \mathbb{P}^{\mu \nu} \mathbb{P}^{\rho \sigma}k_\nu k_\sigma ) \int d^{d-2}\Omega ,\\
\int d^{d-2} \Omega \frac{q^\mu q^\rho q^\lambda}{(P_H-2q)\cdot k}
&=
\int d^{d-2} \Omega \frac{- q^\mu q^\rho q^\lambda}{(P_H+2q)\cdot k}
\nonumber\\
&=
\biggr( \mathcal {F}_3 ( \mathbb{P}^{\mu \rho} \mathbb{P}^{\lambda \sigma}k_\sigma + \mathbb{P}^{\mu \lambda} \mathbb{P}^{\rho \sigma}k_\sigma + \mathbb{P}^{ \rho \lambda} \mathbb{P}^{\mu \sigma}k_\sigma)
\nonumber\\
&~~~ + \mathcal {G}_3 \mathbb{P}^{ \mu \nu} \mathbb{P}^{ \rho \sigma} \mathbb{P}^{ \lambda \tau} k_\nu k_\sigma k_\tau \biggr) \int d^{d-2} \Omega .
\end{align}
\end{subequations}
Where
\begin{subequations}
\begin{align}
\mathcal {T} =& \,_2F_1\left( \frac{1}{2} , 1 , \frac{3}{2} - \epsilon , \Delta^2  \right), \\
\mathcal {F}_1 =& \frac{P_H^2(\mathcal {T}  -1)}{2 (P_H\cdot k)^2},\\
\mathcal {F}_2 =& -\frac{1}{d-2} \frac{P_H^2(\mathcal {T} - 1) + 4q^2 \mathcal {T}}{4 P_H\cdot k } , \\
\mathcal {G}_2 =&  \frac{4q^2 P_H^2 \mathcal {T} + (d-1) (\mathcal {T} - 1)P_H^4}{4(d-2) (P_H\cdot k )^3 }, \\
\mathcal {F}_3 =&    -\frac{P_H^4}{2 (d-2) (P_H\cdot k)^2} \biggr [\biggr (  \frac{q^2}{P_H^2} + \frac{1}{4}  \biggr )(  \mathcal {T} -1 ) + \frac{1}{d-1} \frac{q^2}{P_H^2} \biggr], \\
\mathcal {G}_3 =&   \frac{d+1}{d-2} \frac{P_H^6}{2 (P_H\cdot k)^4} \biggr [\biggr ( \frac{3}{d+1} \frac{q^2}{P_H^2} + \frac{1}{4}  \biggr )( \mathcal {T} -1 ) + \frac{1}{d-1} \frac{q^2}{P_H^2} \biggr].
\end{align}
\end{subequations}

\section{Finite results} \label{sec:app-finite-function}
In this appendix, we provide the expressions of functions
$\mathcal {A}(\Delta)$, $\mathcal {B}(\Delta)$, $\mathcal {C}(\Delta)$, $\mathcal {R}(z,M_H,\Delta)$, $\mathcal{F}(z,M_H,\Delta,\mu)$ and the NLO hard part in Sec.~\ref{sec:perturbative-FF}:
\begin{align}
\mathcal {A}(\Delta)
=&
\frac{1}{4 {\Delta} \left(\sqrt{1-{\Delta}^2}+1\right) \left(\sqrt{1-{\Delta}^2}+2\right)^4}\nonumber\\
&\times
\bigg(\left({\Delta}^2+1\right) \Big(\left(\sqrt{1-{\Delta}^2}+9\right) {\Delta}^4-2 \left(17 \sqrt{1-{\Delta}^2}+37\right) {\Delta}^2\nonumber\\
&+81 \left(\sqrt{1-{\Delta}^2}+1\right)\Big) \left(8 {P_1}-4
{P_2}+4 \ln \left(4 \left(1-{\Delta}^2\right)\right) \tanh ^{-1}({\Delta})+4 \pi ^2\right)\nonumber\\
&-16 \big({\Delta}^6 \left(-\sqrt{1-{\Delta}^2}+\sqrt{1-{\Delta}^2} \ln (2)-10+9 \ln (2)\right)\nonumber\\
&+{\Delta}^4
\left(41 \sqrt{1-{\Delta}^2}+\left(33 \sqrt{1-{\Delta}^2}+65\right) (-\ln (2))+94\right)\nonumber\\
&+{\Delta}^2 \left(-108 \left(\sqrt{1-{\Delta}^2}+1\right)+47 \sqrt{1-{\Delta}^2} \ln (2)+7 \ln (2)\right)\nonumber\\
&+81
\left(\sqrt{1-{\Delta}^2}+1\right) \ln (2)\big) \tanh ^{-1}({\Delta})\bigg).
\end{align}
\begin{align}
\mathcal {B}(\Delta)
=&
\frac{1}{2 {\Delta} \left(\sqrt{1-{\Delta}^2}+1\right)
	\left(\sqrt{1-{\Delta}^2}+2\right)^3}\nonumber\\
&\times
\biggl(
\left({\Delta}^2-1\right) \left(-{\Delta}^2+4 \sqrt{1-{\Delta}^2}+5\right) \left(2 {P_2}-\ln \left(\frac{{\Delta}+1}{1-{\Delta}}\right) \ln \left(-\frac{4}{{\Delta}^2-1}\right)\right)\nonumber\\
&+8 \sqrt{1-{\Delta}^2}
{\Delta}^3 \ln \left(\frac{1}{4} \left(1-{\Delta}^2\right)\right)+8 {\Delta} \Big({\Delta}^4+36 \left(\sqrt{1-{\Delta}^2}+1\right)\nonumber\\
&-{\Delta}^2 \left(8 \sqrt{1-{\Delta}^2}+25+9\ln (2)\right)+\left(14
\sqrt{1-{\Delta}^2}+13\right) \ln (2)\Big)\nonumber\\
&+4 \left(9 {\Delta}^2-14 \sqrt{1-{\Delta}^2}-13\right) {\Delta} \ln \left(1-{\Delta}^2\right)
\biggr).
\end{align}
\begin{align}
&\mathcal {C}(\Delta)\nonumber\\
=&
\frac{1}{36 {\Delta} \left(\sqrt{1-{\Delta}^2}+1\right) \left(\sqrt{1-{\Delta}^2}+2\right)^4}\nonumber\\
&\times
\biggl(
-36 {P_1} \left(3 \left(\sqrt{1-{\Delta}^2}+8\right) {\Delta}^4-2 \left(41 \sqrt{1-{\Delta}^2}+84\right) {\Delta}^2+175 \sqrt{1-{\Delta}^2}+176\right)\nonumber\\
&+18 {P_2}
\left(\left(\sqrt{1-{\Delta}^2}+9\right) {\Delta}^4-2 \left(18 \sqrt{1-{\Delta}^2}+43\right) {\Delta}^2+107 \sqrt{1-{\Delta}^2}+109\right)\nonumber\\
&+{\Delta} \Big(33 \pi ^2 \left(\left(\sqrt{1-{\Delta}^2}+9\right) {\Delta}^4-2
\left(17 \sqrt{1-{\Delta}^2}+37\right) {\Delta}^2+81 \left(\sqrt{1-{\Delta}^2}+1\right)\right)\nonumber\\
&-8 \left(\left(47 \sqrt{1-{\Delta}^2}+369\right) {\Delta}^4-\left(1211 \sqrt{1-{\Delta}^2}+2371\right) {\Delta}^2+2340
\sqrt{1-{\Delta}^2}+2358\right)\Big)\nonumber\\
&-18 \tanh ^{-1}({\Delta}) \Big(-4 \Big(\left(3 \sqrt{1-{\Delta}^2}+22\right) {\Delta}^4-4 \left(16 \sqrt{1-{\Delta}^2}+27\right) {\Delta}^2\nonumber\\
&+85
\sqrt{1-{\Delta}^2}+86\Big)-\Big(5 \left(\sqrt{1-{\Delta}^2}+9\right) {\Delta}^4-2 \left(86 \sqrt{1-{\Delta}^2}+191\right) {\Delta}^2\nonumber\\
&+431 \sqrt{1-{\Delta}^2}+433\Big) \ln
\left(-\frac{4}{{\Delta}^2-1}\right)+8 {\Delta} \bigg(\left(\sqrt{1-{\Delta}^2}+9\right) {\Delta}^4\nonumber\\
&-2 \left(17 \sqrt{1-{\Delta}^2}+37\right) {\Delta}^2+81 \left(\sqrt{1-{\Delta}^2}+1\right)\bigg) \tanh
^{-1}({\Delta})\Big)
\biggr).
\end{align}
\begin{align}
&\mathcal {R}(z,M_H,\Delta)\nonumber\\
=&
-\frac{\delta(1-z)}{96 \left(M_H+m_Q\right)^3 \left(M_H+2m_Q\right)^2}
\Bigl(
16 m_Q^4 \left(3 \mathcal {T} (29 \mathcal {T}-46)+4 \pi^2+60\right) M_H\nonumber\\&
+4 m_Q^3 \left(-114 \mathcal {T}+25 \pi ^2+96\right) M_H^2+4 _Q^2 \left(-90 \mathcal {T}+19 \pi ^2+66\right)M_H^3
\nonumber\\&
+ \left(15+28 \pi ^2\right) m_Q M_H^4+\left(15+4 \pi ^2\right) M_H^5+16 m_Q^5 \left(3 \mathcal {T} (5 \mathcal {T}-8)+\pi ^2+12\right)
\Bigr)\nonumber\\
&+\frac{1}{3 \left(M_H+m_Q\right)^2 \left(M_H+2m_Q\right)^2}
\bigg(3 m_Q^3 M_H \Big(9 \mathcal {T}^2 (z-3) (z-1) z +z (z (20 z-27)+3)
\nonumber\\&
+ \mathcal {T} (6-3 (z-1) z(13 z-14)) -12\Big)+\frac{1}{4} m_Q^2 M_H^2 \big(27 \mathcal {T}^2
(z-1) z (2 z-3)
\nonumber\\&
-54 \mathcal {T} (z-1) (z (5 z-2)+2)+z (7 z (8 z+9)-159)-120\big)
\nonumber\\
& -\frac{27}{2} m_Q (\mathcal {T}-1) (z-1) \ln (1-z) \left(\frac{M_H}{2}+m_Q\right) \Big(2 m_Q M_H (\mathcal {T}
(-z)+\mathcal {T}+z+1)\nonumber\\
&+M_H^2+4 m_Q^2 (-\mathcal {T} (z-2)+z-1)\Big)-\frac{3}{4} m_Q M_H^3 \Big(3 \mathcal {T} (z-1) \left(3 z^2+4\right)\nonumber\\
&+z (z (17 z-45)+48)+6\Big)+\frac{1}{16} (z (13 (9-4 z)
z-141)+24) M_H^4\nonumber\\
& +m_Q^4 \Big(z \big(36 \mathcal {T}^2 ((z-3) z+3)-18 \mathcal {T} (z (5 z-12)+9)+z (50 z-99)+57\big)-12\Big)\bigg)
\nonumber\\
& \times \biggr[ \frac{1}{1-z}\biggr]_+
-\frac{2  ((z-1) z+1)^2}{z}
\biggr[ \frac{\ln(1-z)}{1-z}\biggr]_+.
\end{align}
\begin{align}
&\mathcal
{F}(z,\Delta,M_H,\mu)\nonumber\\
=&
\mathcal {R}(z,M_H,\Delta)N_c-
\frac{1}{36 \Delta  N_c \left(M_H+m_Q\right){}^2
	\left(M_H+2 m_Q\right)^2}\delta(1-z)\nonumber\\
&\times
\biggl(
\left(M_H+m_Q\right){}^2 \left(M_H+2 m_Q\right){}^2 \bigg(\Delta  \Big(9 \mathcal {A}(\Delta)+N_c \big(20 n_f-N_c (9 \mathcal {B}(\Delta)+18
\mathcal {C}(\Delta)\nonumber\\
&-21 \pi ^2+62)\big)+18 \ln \left(1-\Delta ^2\right)+12 \left((3 P_3+8) N_c^2-2 n_f N_c +3\right) \ln
\left(M_H\right)\nonumber\\
&+6 N_c \ln \left(\mu ^2\right) (2 n_f-11 N_c)-36 (1+\ln (2))\Big)\nonumber\\
&-18 \left(1+\Delta ^2\right) \tanh ^{-1}({\Delta}) \left(\ln \left(\frac{1-\Delta^2}{4}\right)+2 \ln \left(M_H\right)\right)\bigg)\nonumber\\
&+480 \Delta  N_c^2 M_H^2 m_Q^2 \ln
\left(M_H\right)+234 \Delta  N_c^2 M_H^3 m_Q \ln \left(M_H\right)\nonumber\\
&+48 \Delta  N_c^2 (1-9 (\mathcal {T}-1) \mathcal {T}) m_Q^4 \ln \left(M_H\right)+72 \Delta
N_c^2 (8-3 \mathcal {T}) M_H m_Q^3 \ln \left(M_H\right)\nonumber\\
&+39 \Delta  N_c^2 M_H^4 \ln \left(M_H\right)
\biggr)
-\frac{\left(z^2-z+1\right)^2 }{z}
N_c
\ln\left(
\frac{M_H^2}{\mu^2}
\right)
\biggr[ \frac{1}{1-z}\biggr]_+
.
\end{align}
\begin{align}\label{eq:NLOhard}
&\hat{D}_{[SS]}^{NLO}(z, M_H, m_Q, \mu, \mu_f=\mu)\nonumber\\
=&
\frac{32\alpha_s^2 \left(M_H+m_Q\right){}^2}{9 \left(N_c^2-1\right) M_H^5}
\Bigg\{
\mathcal{F}(z,\Delta,M_H,\mu)
+\delta(1-z)
\Biggl[
\frac{-\ln(4\pi\mu^2e^{-\gamma_E})}{24 \Delta ^2 N_c \left(M_H+m_Q\right)^2 \left(M_H+2 m_Q\right)^2}\nonumber\\
&\times
\bigg(
12 \Delta  \left(\left(1+\Delta ^2\right) \tanh ^{-1}(\Delta )- \Delta \right) \left(M_H+m_Q\right)^2 \left(M_H+2
m_Q\right){}^2\nonumber\\
&-N_c^2 \Big(4 \Delta^2 (39 P_3+1) M_H^2 m_Q^2+6 \Delta ^2 (12 P_3+1) M_H^3 m_Q\nonumber\\
&+24 \Delta  M_H m_Q^3 \left(2 \Delta -3 \tanh
^{-1}(\Delta )+6 \Delta  P_3\right)+\Delta^2 (12 P_3+1) M_H^4\nonumber\\
&+16 m_Q^4 \left(9 \left(\Delta -\tanh ^{-1}(\Delta )\right) \tanh ^{-1}(\Delta
)+\Delta ^2 (3 P_3-2)\right)\Big)
\bigg)\nonumber\\
&
+\frac{1}{288 \Delta ^2 \left(\sqrt{1-\Delta ^2}+1\right)^2 \left(\sqrt{1-\Delta^2}+2\right)^3 N_c}
\Biggl(
24 \Delta\tanh ^{-1}(\Delta )\nonumber\\
&\times\Bigg(18 \left(1-\Delta ^2\right) \left(\left(\sqrt{1-\Delta^2}+1\right) \Delta ^2+\sqrt{1-\Delta ^2}-1\right) N_c^2 \ln \left(\frac{\mu ^2}{M_H^2}\right)\nonumber\\
&+6 \ln(4\pi e^{-\gamma_E})\bigg(\left(\Delta ^2+1\right) \Big(\left(\sqrt{1-\Delta ^2}+8\right) \Delta^4-27 \left(\sqrt{1-\Delta ^2}+2\right) \Delta ^2\nonumber\\
&+54\left(\sqrt{1-\Delta^2}+1\right)\Big)-2 \Big(3 \left(\sqrt{1-\Delta ^2}+4\right) \Delta^4-32 \left(\sqrt{1-\Delta ^2}+2\right)\Delta ^2\nonumber\\
&+57 \sqrt{1-\Delta ^2}+60\Big) N_c^2\bigg)+N_c^2\left(1+\Delta ^2\right) \Big(2 \Delta ^2 \left(4 \sqrt{1-\Delta^2}+65\right)\nonumber\\
&-3 \left(23 \sqrt{1-\Delta ^2}+76\right)\Big) \Bigg)+4 \Delta  \Bigg(6 \Delta N_c^2 \ln\left(\frac{\mu ^2}{M_H^2}\right) \bigg(3 \Big(\left(\sqrt{1-\Delta ^2}+8\right) \Delta ^4\nonumber\\
&-27 \left(\sqrt{1-\Delta ^2}+2\right)\Delta ^2+54 \left(\sqrt{1-\Delta ^2}+1\right)\Big) \ln \left(\frac{\mu ^2}{M_H^2}\right)\nonumber\\
&-2 \left(\left(\sqrt{1-\Delta^2}+26\right) \Delta ^4-9 \left(10 \sqrt{1-\Delta ^2}+19\right) \Delta ^2+9 \left(20 \sqrt{1-\Delta^2}+19\right)\right)\bigg)\nonumber\\
&+\Delta  \bigg(12 \ln(4\pi e^{-\gamma_E}) \bigg(\Delta ^4 \left(2 \left(5 \sqrt{1-\Delta ^2}+28\right) N_c^2-3\left(\sqrt{1-\Delta ^2}+8\right)\right)\nonumber\\
&-3 \left(\sqrt{1-\Delta ^2}+2\right) \Delta ^2 \left(58 N_c^2-27\right)+9\sqrt{1-\Delta ^2} \left(37 N_c^2-18\right)\bigg)\nonumber\\
&+\left(4 \left(5 \sqrt{1-\Delta ^2}+82\right) \Delta ^4-12 \left(28\sqrt{1-\Delta ^2}+89\right) \Delta ^2+369 \sqrt{1-\Delta ^2}\right) N_c^2\bigg)\nonumber\\
&+18 \Delta  \left((228 \ln(4\pi e^{-\gamma_E})+47)N_c^2-108 \ln(4\pi e^{-\gamma_E})\right)+3 \pi ^2 \Big(\left(\sqrt{1-\Delta ^2}+8\right) \Delta ^4\nonumber\\
&-27 \left(\sqrt{1-\Delta ^2}+2\right)\Delta ^2+54 \left(\sqrt{1-\Delta ^2}+1\right)\Big) \left(\Delta  \left(6 \Delta +N_c^2\right)+6\right)\Bigg)\nonumber\\
&+9\left(\Delta ^2-1\right)^2 N_c^2 \ln ^2\left(\frac{1+\Delta }{1-\Delta }\right) \bigg(5 \sqrt{1-\Delta ^2}+12\left(\sqrt{1-\Delta ^2}+2\right)
\ln\left(\frac{4\pi\mu^2e^{-\gamma_E}}{M_H^2}\right)\nonumber\\
&+58\bigg)
\Biggr)
\Biggr] +\frac{2 z^4 N_c \left(M_H+z m_Q\right)^2}{\left(M_H+m_Q\right)^2 }\biggr[\frac{\ln(1-z)}{1-z}\biggr]_+
+
\frac{z^4 N_c \left(M_H+z m_Q\right){}^2}{12  \left(M_H+m_Q\right){}^4 \left(M_H+2 m_Q\right)^2}
\nonumber\\&\times  \biggr[\frac{1}{1-z}\biggr]_+
\Biggl[
-16 m_Q^4 \bigg(3 \ln \left(\frac{\mu ^2 z^2}{M_H^2}\right)
+\frac{9}{4\Delta^2} \ln ^2\left(\frac{1+\Delta }{1-\Delta }\right)
-9 \frac{\tanh ^{-1}(\Delta )}{\Delta }-1\bigg)
\nonumber\\
&-24 M_H m_Q^3 \left(6 \ln \left(\frac{\mu ^2 z^2}{M_H^2}\right)+3 \frac{\tanh ^{-1}(\Delta )}{\Delta }-8\right)-4 M_H^2 m_Q^2 \left(39
\ln \left(\frac{\mu ^2 z^2}{M_H^2}\right)- 40\right)\nonumber\\
&-(6 M_H^3 m_Q+M_H^4) \left(12 \ln \left(\frac{\mu ^2
	z^2}{M_H^2}\right)- 13\right)\Biggr]
\Bigg\}.
\end{align}
where $\tanh^{-1}({\Delta})
=
\frac{1}{2}(\ln
\left({1+{\Delta}}\right)-\ln
\left({1-{\Delta}}\right))$ and we defined:
\begin{subequations}
	\begin{align}
	P_1
	=&
	\text{Li}_2(-\Delta)-\text{Li}_2(\Delta),\\
	P_2
	=&
	\text{Li}_2\left(\frac{1-\Delta}{2}\right)-\text{Li}_2\left(\frac{1+\Delta}{2}\right),\\
	P_3
	=&
	\frac{1}{3 \Delta ^3 \left(\Delta ^2+3\right)}
	\biggl(\left(9 \Delta ^4+\left(9 \sqrt{1-\Delta ^2}+6\right) \Delta ^2-9 \sqrt{1-\Delta ^2}+9\right) \tanh ^{-1}(\Delta )\nonumber\\
	&-\Delta  \left(8 \Delta ^4+\left(6
	\sqrt{1-\Delta ^2}+9\right) \Delta ^2-9 \sqrt{1-\Delta ^2}+9\right)\biggr).
	\end{align}
\end{subequations}

\input{ref.bbl}

\end{document}

%% file: ref.bbl
\providecommand{\href}[2]{#2}\begingroup\raggedright\endgroup